\documentclass{jfm}

\usepackage{natbib}
\usepackage{latexsym}
\usepackage{amsmath}
\usepackage{upmath}
\usepackage{amsbsy}
\usepackage{amssymb}
\usepackage{amsgen}
\usepackage{amsfonts}
\usepackage{hyperref}
\usepackage{graphicx}
\usepackage{array}
\usepackage{ulem}
\usepackage{color}
\usepackage{cases}
\usepackage{fancyvrb}
\usepackage[english]{babel}
\usepackage[utf8]{inputenc}
\usepackage{lmodern}
\usepackage{rotating}
\usepackage{hhline}
\usepackage[T1]{fontenc}
 
\newcommand{\gerris}{{\usefont{T1}{pzc}{m}{n}Gerris}}
\title{Drop impact on a solid surface : short time self-similarity}
\author[J. Philippi, P.-Y. Lagr\'ee and A. Antkowiak] 
{J\ls U\ls L\ls I\ls E\ls N\ns P\ls H\ls I\ls L\ls I\ls P\ls P\ls I,
\ns P\ls I\ls E\ls R\ls R\ls E\ls -\ls Y\ls V\ls E\ls S\ns  L\ls A\ls G\ls R\ls \'E\ls E\ns\and
\ns A\ls R\ls N\ls A\ls U\ls D\ns A\ls N\ls T\ls K\ls O\ls W\ls I\ls A\ls K\ns
}
\affiliation{Sorbonne Universit\'es, UPMC Univ Paris 06, CNRS, UMR 7190 Institut Jean Le Rond d'Alembert, F-75005 Paris, France.}

\begin{document}

\maketitle

\begin{abstract}
The early stages of drop impact onto a solid surface are considered. Detailed numerical simulations and detailed asymptotic analysis
of the process reveal a self-similar structure both for the velocity field and the pressure field. The latter is shown to exhibit a maximum not near the impact point, but rather at the contact line. The motion of the contact line is furthermore shown to exhibit a 'tank treading' motion. These observations are apprehended at the light of a variant of Wagner theory for liquid impact. This framework offers a simple analogy where the fluid motion within the impacting drop may be viewed as the flow induced by a flat rising expanding disk. The theoretical predictions are found to be in very close agreement both qualitatively and quantitatively with the numerical observations for about three decades in time. Interestingly the inviscid self-similar impact pressure and velocities are shown to depend solely on the self-similar variables $(r/\sqrt{t},z/\sqrt{t})$.
The structure of the boundary layer developing along the wet substrate is investigated as well, and is proven to be formally analogous to that of the boundary layer growing in the trail of a shockwave. Interestingly, the boundary layer structure only depends on the impact self-similar variables.
This allows to construct a seamless uniform analytical solution encompassing both impact and viscous effects.
The depiction of the different dynamical fields enables to quantitatively predict observables of interest, such as the evolution of the integral viscous shearing force and of the net normal force.
\end{abstract}

%
%

\section{Introduction}
\label{sec:introduction}

The impact of a liquid drop onto a rigid surface results in a rapid sequence of events ending, in the inertial limit, in spreading \citep{Eggers2010a} or splashing \citep{Stow1981}, interface tearing \citep{Villermaux2011} and ultimate fragmentation \citep{Stow1977}. A large number of studies have investigated the many facets of drop impact, with a special attention to the description of its late stages \citep{Rein1993,Yarin1995}. The literature on the early stages of impact is however scarce in comparison. Detailed experimental data depicting the instants following impact can nonetheless be found in the work of \citet{Rioboo2002}, that evidenced a ``kinematic phase'' where the drop merely resembles a truncated sphere and spreads as the square-root of time. This phase precedes the apparition of the liquid lamella. 

\begin{figure}
\centering
\includegraphics[width=6cm]{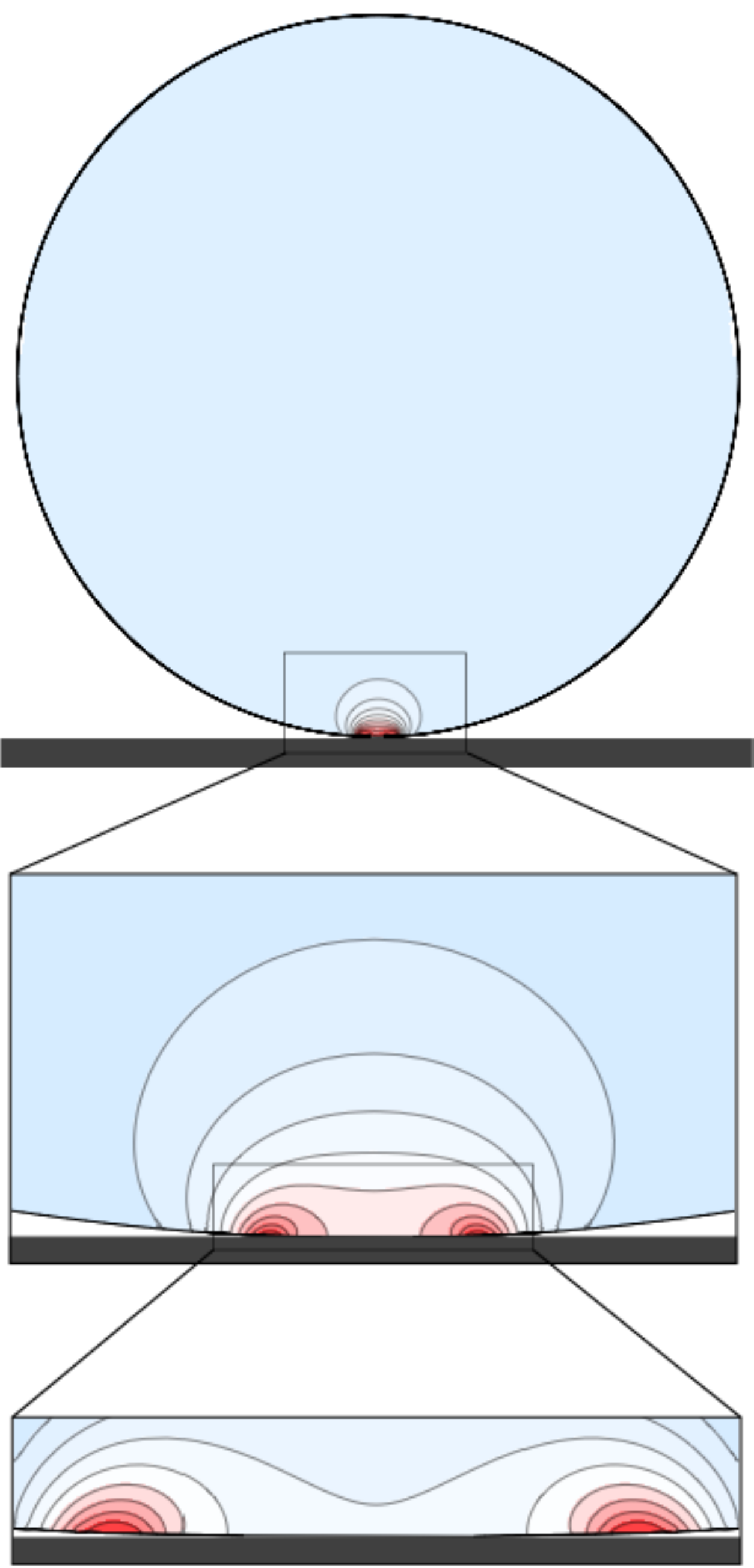}
\caption{Close-ups of increasing magnitude on the pressure field developing inside an impacting drop in the inertial limit. The pressure field is extracted from Navier-Stokes \gerris\ computations of a drop impacting a solid surface at early times (note that the surrounding gas dynamics is computed as well, but not represented here). Noticeably the motion is essentially pressureless (and therefore corresponds to a free fall) except in a concentrated region in the contact zone. The successive close-ups on pressure field structure in the contact region reveal a pressure peak near the contact line (the physical parameters are here Re = 5000, We = 250, $tU/R$=4$\times$10$^{-4}$. The total size of the numerical axisymmetric domain is $2R \times 2R$, and the adaptive mesh has locally a mesh density corresponding to 32768$\times$32768 grid points).} 
\label{fig:tryptic}
\end{figure}

Probably one of the first depiction of the very first instants of drop impact dates back to \citet{Engel1955}. With the help of high-speed cinematography, \citeauthor{Engel1955} captured the chronology of events triggered by drop impact. He noted in particular the unvarying shape of the drop apex during the earliest moments of impact, which might be surprising due to the incompressible character of the liquid. \citeauthor{Engel1955} put forward the possible roles of inertia, viscosity or surface tension to explain this observation. Actually, the physical mechanism underpinning this behaviour is best illustrated with Figure~\ref{fig:tryptic}a. There, the numerically computed pressure field within an impacting drop is represented shortly after impact (details to follow in the paper). It is readily seen that the structure of the pressure field is extremely concentrated near the contact zone, as in Hertz' classic elastic contact problem. Conversely the outer region is essentially pressureless. This strong inhomegeneity in the pressure distribution therefore explains why, in the absence of any pressure hindrance, the upper part of the drop freely falls even after impact while remaining undeformed.

\begin{figure}
\centering
\includegraphics[width=13cm]{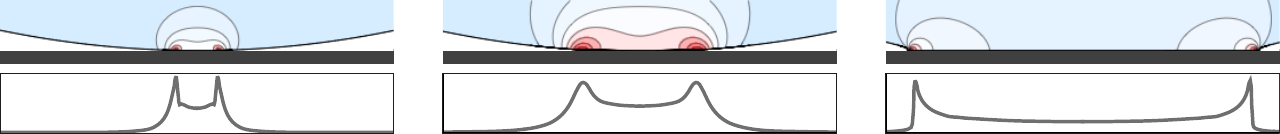}
\caption{Top: Time sequence of the pressure field developing inside an impacting drop (Navier-Stokes {\gerris} computation, fixed spatial magnification). Bottom: Corresponding trace of the pressure exerted by the drop on the solid substrate. The physical parameters for this simulation are $\text{Re} = 5000$ and $\text{We} = 250$. The snapshots correspond respectively to times $tU/R=10^{-4},10^{-3}$ and $10^{-2}$. }
\label{fig:radial_pressure}
\end{figure}

\begin{figure}
\centering
\includegraphics{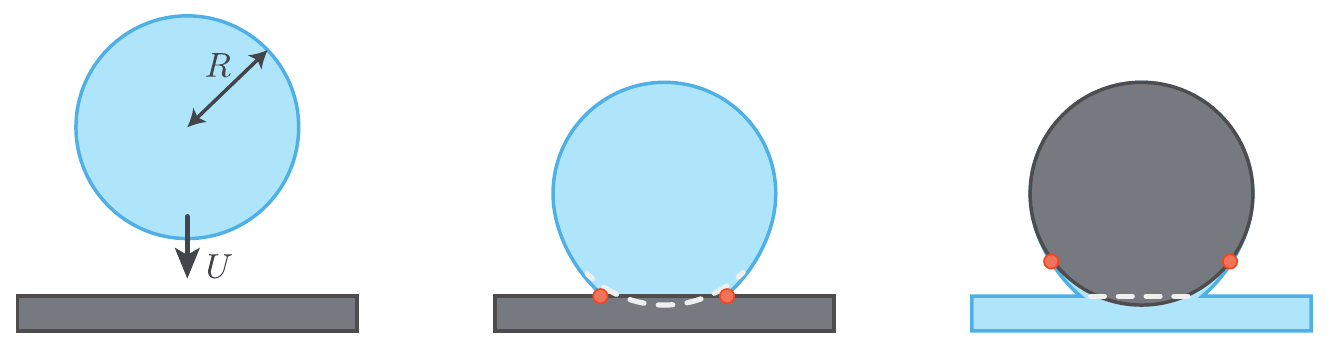}
\caption{Sketch of the impacting drop before contact (left), and shortly after impact (middle). The shape the drop would assume in absence of wall is outlined with a dashed line, and the contact line position is marked with red dots. This problem may be viewed as the dual of the classic water entry of a solid object (right).} 
\label{fig:drop_impact_sketch}
\end{figure}

The pressure concentration in the early stages of impact was first identified by \citet{Josserand2003}. From the key remark that the extent of the pressure concentration zone scales with the contact radius, these authors conjectured a self-similar structure for the pressure field and evolution with time as $1/\sqrt{t}$ -- an hypothesis comforted by numerical results. Though sufficient to detect hints of self-similarity, numerical simulations were nonetheless unable to reveal the inner structure of the contact zone until recently, essentially because of the very large scale ratio between this zone and the drop size. The increase of computational performance along with the development of adaptive numerical techniques for two-phase flows \citep{Popinet2009} now allow to unravel the intimate structure of the contact zone, see Figures~\ref{fig:tryptic}b,c and~\ref{fig:radial_pressure}. These snapshots reveal a quite complex structure for the pressure, which counter-intuitively exhibits sharp maxima near the contact line, and not on the axis as in steady stagnation point flows.
Interestingly this structure is reminiscent of typical pressure field structures observed in the water entry of solid objects, and evidenced by Wagner in the context of alighting seaplanes \citep{Wagner1932}.
In such problems a solid object impacts a flat liquid surface at a given velocity. Drop impact may be viewed as water entry's opposite, for a liquid object impacts a rigid flat surface at a given velocity (see Fig.~\ref{fig:drop_impact_sketch}).
It is therefore likely that the analytical techniques developed since the thirties to describe with great precision the flow generated with the impact of an object, and proven to be in close agreement with experimental data \citep{Howison1991}, could be transposed for the drop impact problem. And indeed, building up on the analogy between water entry and drop impact, \citet{Howison2005} proposed a theoretical investigation of two-dimensional drop impact on a thin fluid layer and described the different regions and scalings of importance for the flow dynamics. In particular, they reveal the radius of contact between the two liquids as a key length scaling the problem, analogously to the problem of water entry where the wet length of the solid is also determining and consistently with the observations of \citet{Josserand2003}.

The central motivation of the present study is to revisit the problem of a single spherical drop impacting a smooth flat solid surface at early times at the light of Wagner's theory of impact, understand the dynamic fields structure and elucidate the short-time self-similar behaviour discerned in earlier studies. To develop a consistent theory, the approach followed throughout the manuscript will be to confront and cross-test systematically the theoretical predictions with detailed and accurate numerical simulations performed with a Navier-Stokes multiphase flow solver. As a side note, we essay to make the paper self-contained whenever possible.
In \S 2 we formulate the hypotheses and theoretical framework of the problem, and describe the short-time drop impact dynamics in the context of Wagner's theory. We put forward in particular a so-called ``Lamb analogy'' mirroring the flow within the impacting drop with the one induced with a flat rising expanding disk. In \S 3 we demonstrate that the Wagner flow can be recast as a self-similar solution for the drop impact problem. The nature of the near-axis stagnation flow and of the near contact line flow and pressure maxima are also discussed. Numerical results obtained with \gerris\ (Navier-Stokes solver, VOF, adaptive mesh) taking into account surface tension, surrounding gas and viscous effects are compared with the theoretical prediction.  The structure of the boundary layer is examined in \S 4 and is found to be reminiscent of the viscous boundary layer leaved in the trail of a shockwave (Mirels analogy). The inviscid Wagner flow and this viscous boundary layer are found to depend on the same self-similar variable. This allows us to build a uniform, seamless solution encompassing both impact and viscous effects. We conclude in \S 5 by discussing the obtained results and the limits of the present investigation (such as the role of air), and by reminding observables of interest, such as the net impacting force of the total viscous shearing force exerted by an impacting drop.

%
%

\section{Model}
\label{sec:model}
\subsection{Theoretical framework \& hypotheses}
\label{sec:framework}
We consider throughout this study an idealized drop impact where a perfectly spherical liquid drop collides with a flat rigid surface.
Though classic, this model situation relies on a number of physical hypotheses detailed in the following. Starting with the initial perfect spherical shape assumption, we may identify several typical causes for deviations from sphericity such as capillary drop oscillations during free fall \citep{Engel1955,Thoroddsen2005} or flowing air shaping \citep{Pruppacher1970}. Such effects will be disregarded in the following even though they merely result in a local curvature radius modification in the contact region, hence could easily be incorporated in the discussion. 
The impact flow evolution will be considered incompressible. However, the question of the role of compressible effects within the liquid should not be eluded when considering impact phenomena. A large body of literature has been devoted to acoustic effects within (typically fast) impacting drops. Typically, such effects are considered to play a significant role over lengthscales of order $U R/c$ during timescales of order $U R / c^2$, where $U$ stands for the impact velocity, $R$ for the drop radius and $c$ for the celerity of compressive sound waves in the liquid ($(\frac{U}{c})^2 \ll \frac{U t}{R} \ll 1$). Considering a millimetric drop impacting at a velocity of the order of 10 m.s$^{-1}$ representative of \textit{e.g.} a raindrop falling at terminal velocity, we reckon that acoustic effects matter only in a micron-sized region over a few nanoseconds \citep[see][for a discussion and references]{Weiss1999}. The following discussion will therefore be limited to those cases where the impact velocity is much lower than the speed of sound, as the falling raindrop, where acoustic effects can harmlessly be neglected and an incompressible description remains accurate.
The high pressure and stresses generated upon impact can result in marked erosion or yielding \citep{Rein1993}. Furthermore substrate deformation has recently been shown to significantly alter drop impact in the limit of very soft \citep{Mangili2012} or very flexible substrates \citep{Antkowiak2011}. None of these effects will be considered in the following, yet an estimation of the net force exerted by the impacting drop on the underlying substrate will be provided in \S \ref{sec:selfsimilarsolution}.
Another phenomenon potentially responsible for the cushioning of the impact is the thin air layer between the drop and the substrate just before impact. Due to lubrication effects, this air layer pressurizes and dimples the drop, eventually yielding a tiny entrapped gas bubble in the drop \citep{Thoroddsen2005}. This phenomenon along with other roles of surrounding gas in impact dynamics will be disregarded in the forthcoming analysis (except explicitly specified, the starting state for each of the simulations is a drop already touching the ground -- see appendix for details) and discussed in the last section.
Finally the core hypothesis of the present study is the inertia-dominated character of impact. In particular, we assume that gravity, capillary and viscous effects are small with respect to inertial ones, \textit{i.e.} Froude $\text{Fr}=U^2/g R$, Weber $\text{We} = \rho U^2 R/\sigma$ and Reynolds $\text{Re} = \rho U R/\mu$ numbers are all large with respect to unity. Here $g$ denotes the gravity, $\sigma$ the liquid-gas surface tension, $\rho$ the liquid density and $\mu$ its viscosity. 
These assumptions underpin the choice a purely inertial description free of these effects in the following. However, locally these phenomena might become more important or even dominant, examples being viscosity near the boundaries or capillarity in high-curvature region. In section \S 4 we will address viscous effects and develop a boundary-layer correction to the inviscid solution, and capillary effects will eventually be discussed in \S 5.
\subsection{Governing equations and analogy with the water entry problem}
\label{sec:governing_equations}
\subsubsection{Problem statement}
We consider a perfectly spherical liquid drop of radius $R$ and density $\rho$ impacting normally a flat rigid ground with velocity $U$, see Fig.~\ref{fig:drop_impact_sketch}. Neglecting for now the development of viscous rotational boundary layers, we assume that the fluid motion following impact is irrotational, axisymmetric and can be described with the scalar potential $\phi(r,z)$, \textit{i.e.} the fluid velocity $\boldsymbol{u}(r,z)$ satisfies  $\boldsymbol{u}(r,z) = \boldsymbol{\nabla} \phi(r,z)$. Incompressibility requires $\phi$ to be an harmonic potential satisfying Laplace's equation, here written in cylindrical coordinates:
\begin{equation}
 \frac{1}{r} \frac{\partial}{\partial r} \left( r \frac{\partial \phi}{\partial r} \right) + \frac{\partial^2 \phi}{\partial z^2} = 0.
 \end{equation}
The liquid dynamics obeys the unsteady form of Bernoulli's conservation equation:
\begin{equation}
\frac{\partial \phi}{\partial t} + \frac{1}{2} \vert \nabla \phi \vert^2 + \frac{p}{\rho} = \mathrm{const}.
\label{eq:unsteady_bernoulli}
\end{equation}
This set of equations is completed by appropriate boundary equations. At the wall $z = 0$, the condition of impermeability reads 
\begin{equation}
\frac{\partial \phi}{\partial z} = 0 \text{ for } 0 \leq r \leq d(t),
\end{equation}
where $d(t)$ stands for the contact line position, an unknown of the problem. The position of the free surface is tracked with the kinematic condition:
\begin{equation}
\frac{\mathrm d\mathcal S}{\mathrm dt} = 0,
\end{equation}
where $\mathcal S(r,z,t)$ is a function vanishing on the free surface.
Expressing normal stress continuity at this interface yields the following dynamic boundary condition:
\begin{equation}
p = 0\text{ at the free surface}.
\end{equation}
Note that atmospheric pressure as here been taken as the reference pressure.

\noindent Anticipating the forthcoming analysis of the contact region, we further note that the free fall behaviour outside the contact region can be recast into the following far-field condition:
\begin{equation}
\phi \to -U z \,\text{ far from the contact region}.
\end{equation}
This condition allows to identify the constant in~(\ref{eq:unsteady_bernoulli}) as $\frac{1}{2} U^2$.

\noindent Now nondimensionalising the problem using the inertial scales $R$, $\rho$ and $U$, we introduce the following quantities: 
\begin{equation}
r = R \, \bar{r},\quad
z = R \, \bar{z},\quad
t=\displaystyle\frac{R}{U} \, \bar{t},\quad
\phi = U R \, \bar{\phi},\quad
p = \rho U^2 \, \bar{p},
\end{equation}
and rewrite the equations into their dimensionless counterparts:
\begin{alignat}{2}
\frac{1}{\bar{r}} \frac{\partial}{\partial \bar{r}} \left( \bar{r} \frac{\partial \bar{\phi}}{\partial \bar{r}} \right) + \frac{\partial^2 \bar{\phi}}{\partial \bar{z}^2} = 0 \quad & \text{in the liquid,} \label{eq:f1}\\
\frac{\partial \bar{\phi}}{\partial \bar{t}} + \frac{1}{2} \vert \bar{\nabla}\bar{\phi} \vert^2 + \bar{p} = \frac{1}{2} \quad & \text{in the liquid,}\\
\frac{\partial \bar{\phi}}{\partial \bar{z}}(\bar{r},\bar{z}=0,\bar{t}) = 0  \quad & \text{over the wet area }\bar{r} < \bar d(\bar{t}), \\
\bar p = 0 \quad & \text{on the free surface,}\\
\frac{\mathrm d \bar{\mathcal S}}{\mathrm d \bar{t}} = 0 \quad & \text{on the free surface.}\label{eq:f5}
\intertext{Finally the nondimensional far-field condition reads:}
\bar{\phi}= -\bar{z} \quad &\text{far from the contact region.}
\end{alignat} 
As posed, the problem entirely depends on the wet area extent $\bar d(\bar t)$, whose dynamics has still to be determined. In the following, we investigate the near-contact line flow to clarify this wetting dynamics.

\begin{figure}
\centering
\includegraphics[width=15pc]{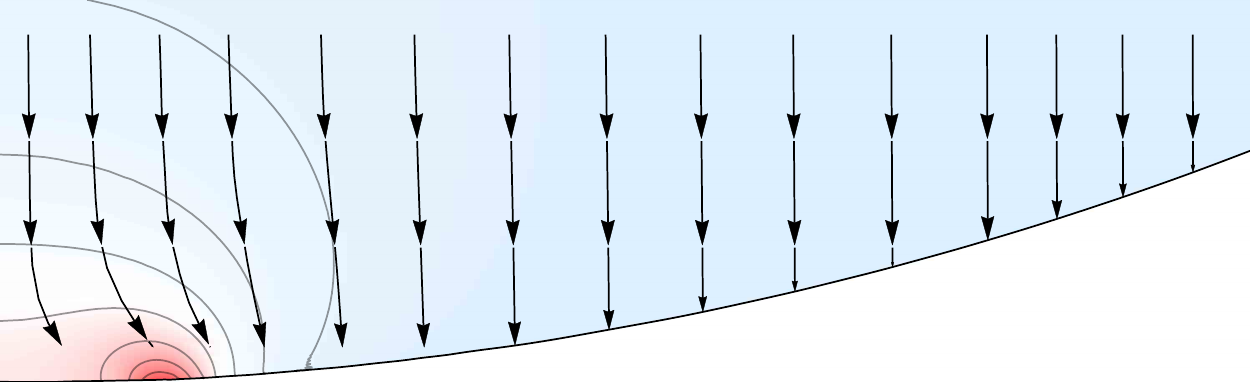}\hspace{2pc}\includegraphics[width=15pc]{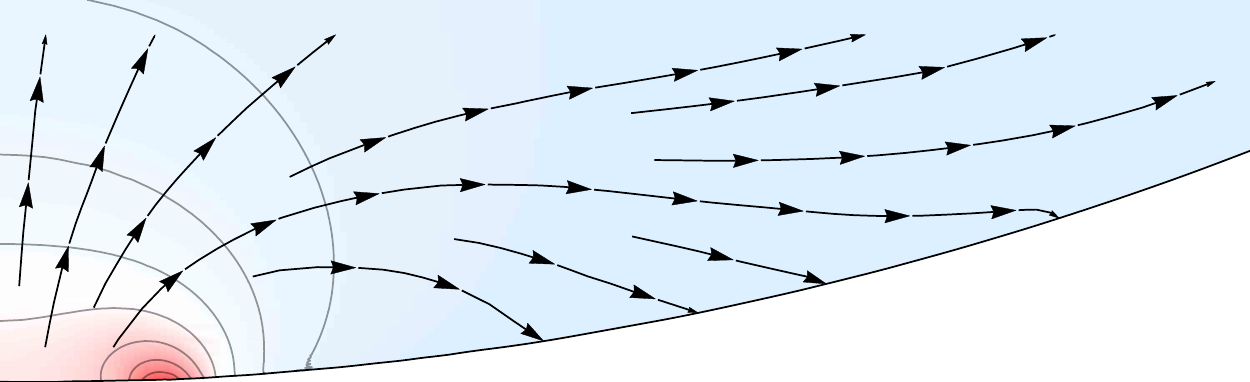}\\
\includegraphics[width=15pc]{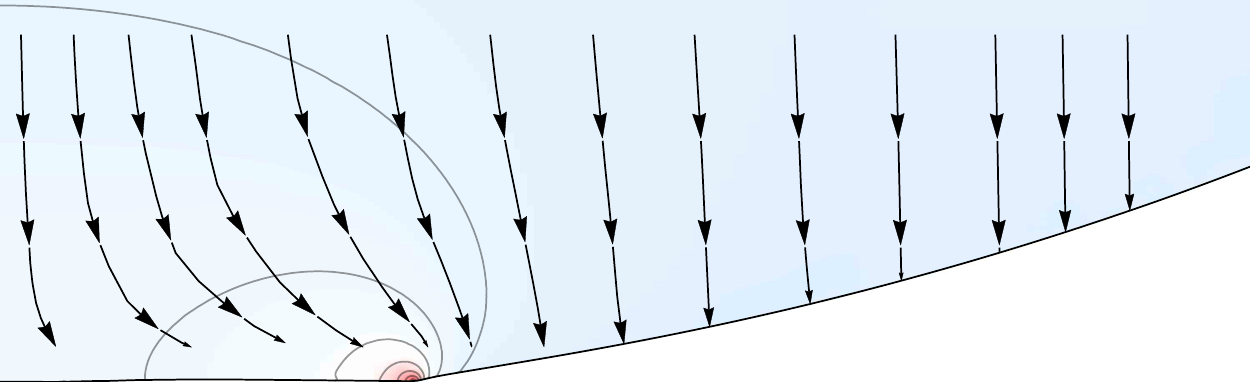}\hspace{2pc}\includegraphics[width=15pc]{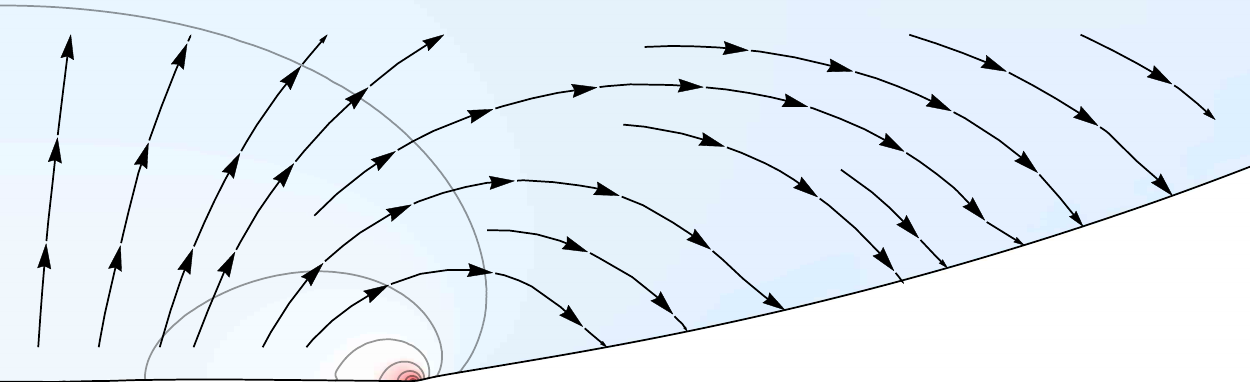}\\
\includegraphics[width=15pc]{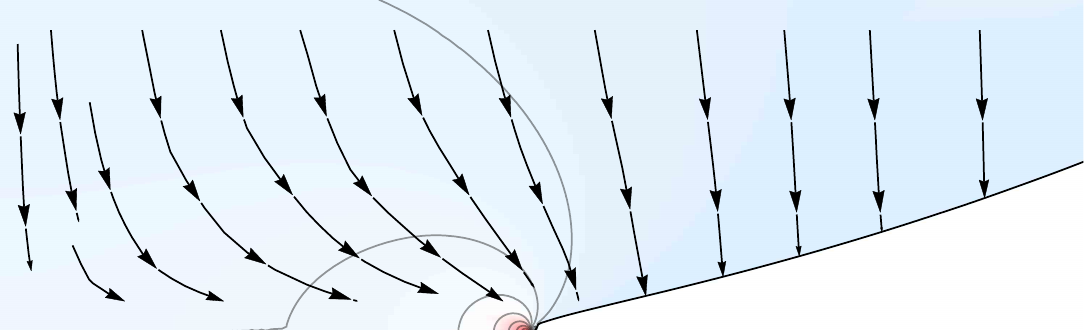}\hspace{2pc}\includegraphics[width=15pc]{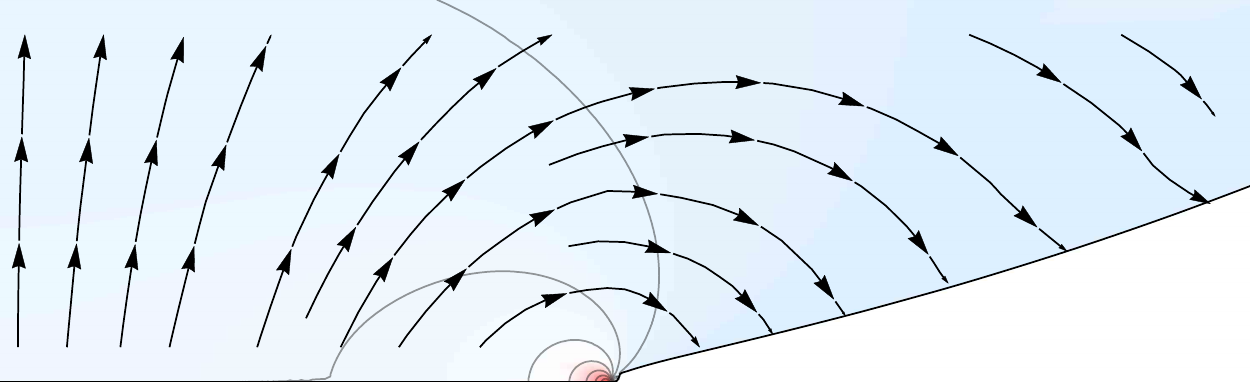}
\caption{Flow in an impacting drop in the fixed frame and in the drop frame computed with \gerris . Left, top to bottom : Streamlines and pressure map within an impacting drop for $\text{Re} = 5000$ and $\text{We} = 250$ in the laboratory frame at different post-impact times ($\bar{t} = 5 \times 10^{-4}$, $5 \times 10^{-3}$ and $10^{-2}$). The overall velocity field resembles a stagnation point flow in a near-wall region whose extent is scaling with the wet area, and a uniform downwards flow outside. 
Right: Same velocity field in a reference frame moving with the drop initial velocity, evidencing a bypass motion near the contact line and an overacceleration of the free surface towards the wall.} 
\label{fig:velocity_field}
\end{figure}
\subsubsection{Contact line motion: numerical observations}
\label{sec:contactlinenumericalobservations}
To shed light on the contact line dynamics, detailed numerical simulations of impacting drops were carried out with \gerris\ (see \S \ref{sec:Annex}). Figure~\ref{fig:velocity_field} represents typical streamlines extracted from the simulations, shortly after impact. On the left panel it can be seen that the motion within the impacting drop far from the contact zone is vertical, uniform and pointing downwards, corresponding merely to the free flight behaviour $-U \boldsymbol{e}_z$. Near the wall though, the flow is deflected and exhibits a stagnation point-like structure, in a region whose extent scales with the wet area. To investigate further the nature of this corrective flow, we represent on the right panel of Fig.~\ref{fig:velocity_field} the streamlines in a reference frame moving with the initial velocity of the drop. There it appears that the flow winds around the contact line, revealing that \textit{(i)} the liquid near the contact line falls faster than free-flight and \textit{(ii)} rather than being pushed by a sweeping motion, the contact line progresses via a tank-treading movement, analogous to the rolling motion evidenced in previous studies of advancing contact lines~\citep{Dussan-V.1974,Chen1997}.

\begin{figure}
\centering
\includegraphics{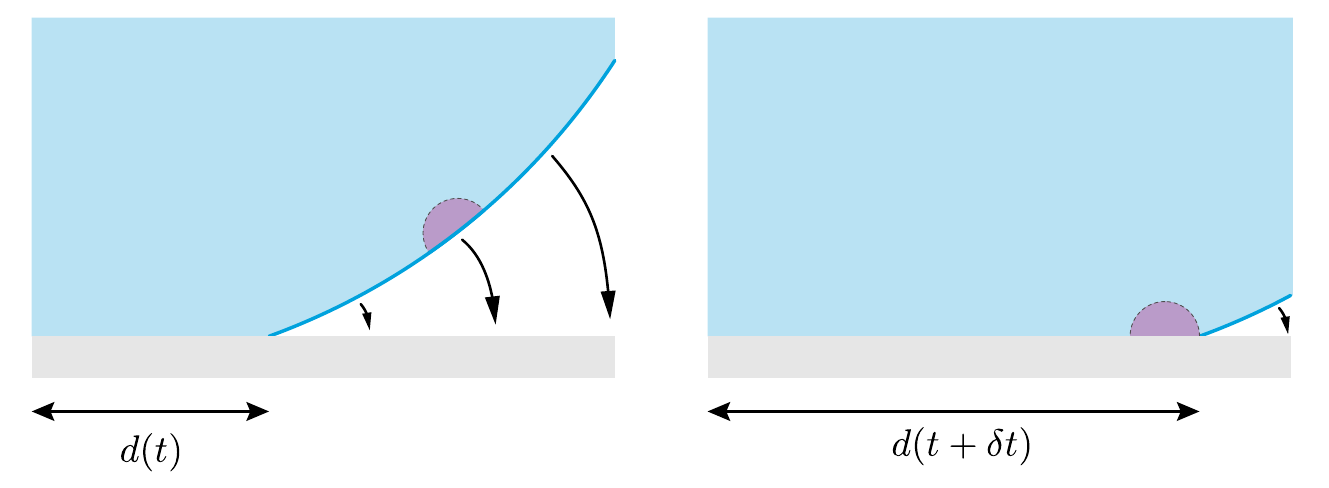}
\caption{Advancing contact line shortly after impact. In the earliest moments following impact, the motion of the free surface near the contact zone is essentially directed downwards. The sketches show the position of the contact line for two successive instants, and illustrate the fact that the horizontal extension of the wet area is governed by the vertical movement of the interface.} 
\label{fig:wagner_condition}
\end{figure}

These observations therefore suggest that the kinematics of horizontal extension for the wet radius is controlled by the vertical motion of the free surface. Figure~\ref{fig:wagner_condition} illustrates this process, and indicates that the law of motion of the contact line $d(t)$ can be obtained from the knowledge of velocity field at the free surface. Such a kinematic condition expressing the contact between a liquid surface and a solid object has actually been used in the context of the water entry of solid objects for about 80 years, and is currently referred to as \textit{Wagner condition}. In the following we depict the analogies between these two liquid impact problems, and use them to derive a simple fluid mechanical model for the drop impact. 

\subsubsection{Analogy with the water entry problem}
\label{sec:analogywaterentry}
The modern understanding of the liquid motion and forces generated by an impacting object in water originates in the pioneering work of Wagner in the early thirties \citep{Wagner1932}. The primary motivation of Wagner was to provide a detailed characterization of the impulsive forces generated with impact -- already known to be of sufficient amplitude to induce bouncing (ricochet), and even possibly structural failure of alighting seaplanes or slammed ships \citep{Nethercote1986}. The foremost issue in this problem evidently stems from its highly unsteady and nonlinear nature. 
The central idea of Wagner was to model the flow induced by the impact of a float or a keel by the one  induced by a flat ``plate'', propelling the fluid particles downwards at the float or keel velocity, and having an extent growing with time as the waterline length. The corresponding flow  (\textsl{``gleiche Tragfl\"ugelbewegung''} -- equivalent aerofoil motion) is typically found to wind around the plate and therefore to promote jetting or splashing. The knowledge of this flow field then allows to determine the motion of the free surface, and finally provides the needed condition in the determination of the wet length $d(t)$.

\begin{figure}
\centering
\includegraphics[width=5cm]{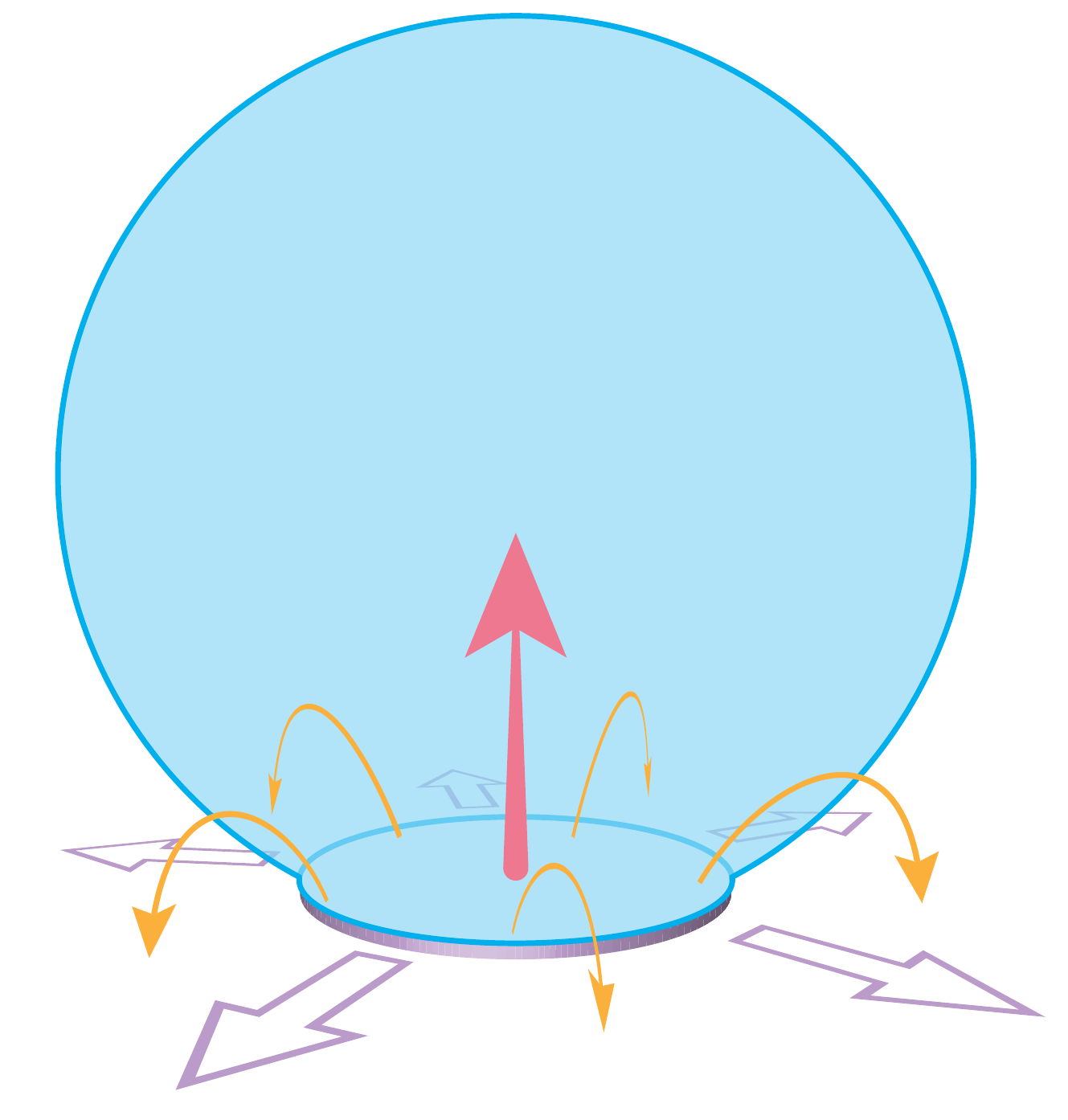}
\caption{In the reference frame of the falling drop, the flow induced by impact may be seen as the one induced by a flat rising disk (Lamb disk analogy). The winding motion is here represented with orange arrows, and the radial expansion of the disk with the wet area is indicated with purple arrows. The motion of the disk itself is given by the red arrow.} 
\label{fig:expanding_plate}
\end{figure}

Analogously, for the drop impact problem, our numerical simulations evidence similar flow features and winding motion. These observations advocate for the use of a water entry-analogue description, where the flow induced by drop impact would correspond to the one induced by a flat expanding disk in normal incidence, which extent is given the wet area (see figure \ref{fig:expanding_plate}). Following this vision of drop impact as a dual version of the water entry problem, we adopt from now on the corresponding formalism to describe the fluid mechanics of impact.

\subsection{Leading-order description for the drop impact problem}
\label{sec:leading-order}
\begin{figure}
\centering
\includegraphics{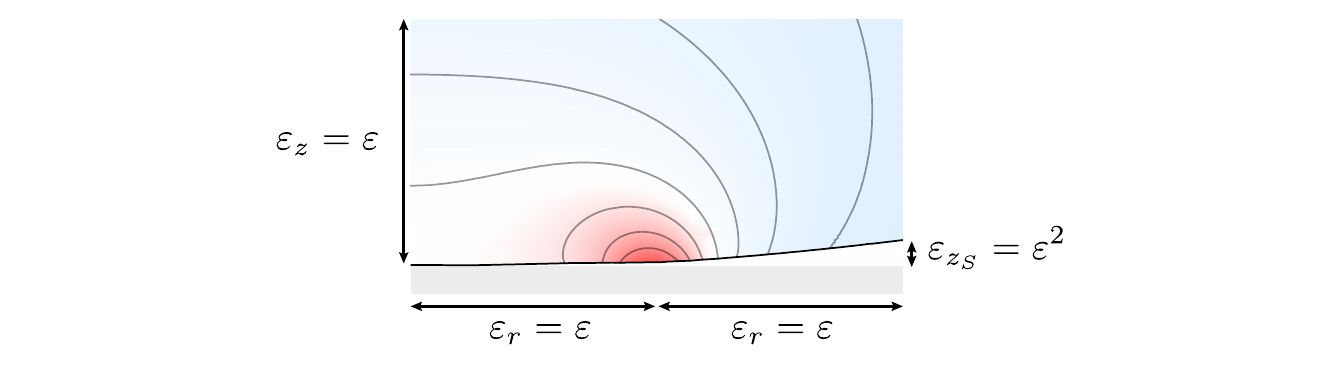}
\caption{Scalings in the contact zone. At the earliest times only a very small portion (of order $\varepsilon$) of the drop touches the wall. The fluid sets into motion with impact is in a region of extent $\varepsilon$ in every direction. The air wedge confined between the wall and the drop presents an angle of order $\varepsilon$ as well. The colormap illustrates the pressure distribution. The physical parameters for this simulation are $\text{Re}=5000$ and $\text{We}=250$. This snapshot corresponds to a nondimensional time $\bar{t}=10^{-3}$. The position of the contact line is here $\bar d=1.73\; 10^{-3/2}$.}
\label{fig:scaling}
\end{figure}

Interested in the early-time behaviour of the impact-induced flow, we set out by examining time-dependent solutions of system (\ref{eq:f1}--\ref{eq:f5}) in the vicinity of the contact zone. To this end, we introduce $\varepsilon$ as a measure of the wet region: $d(t)/R = O(\varepsilon)$ (see Fig.~\ref{fig:scaling}). This $\varepsilon$ is the fundamental small parameter of our problem. 

\noindent As typical in two-phase phenomena, the lengthscales for the dynamical fields and for the geometry of the free surface differ in this problem. Starting by considering the space variables $\bar{r}$ and $\bar{z}$ on which depend the dynamical fields (such as the velocity potential $\phi$ or the pressure $p$), we introduce the following rescaling: $\bar{r} = \varepsilon_r \tilde{r}$ and $\bar{z} =  \varepsilon_z \tilde{z}$, where $\tilde{r}$ and $\tilde{z}$ are $O(1)$ quantities and $\varepsilon_r$ and $\varepsilon_z$ are gauge functions.
From the structure of Laplace operator, we expect the dynamical fields to display identical length scales in each direction, so that
$\varepsilon_r  =  \varepsilon_z = \varepsilon$. 

\noindent Insights into the relevant lengthscales for the description of the free surface geometry can be gained by decomposing the position of the surface into that of a translating sphere $\bar{z}_S(\bar{r},\bar t)$ plus a surface disturbance $\bar h(\bar{r},\bar t)$ (see Fig.~\ref{fig:drop_impact_sketch}b). Assuming the drop falls with constant velocity, the shape of the unperturbed translating sphere obeys $\bar{r}^2 + (\bar{z}_S-(1-\bar{t}))^2 = 1$. 
Sufficiently close to the contact area, we introduce gauge functions for the vertical position of the moving sphere $\bar{z}_S$ and the time $\bar t$: $\bar{z}_S = \varepsilon_{z_S} \tilde z_S$ and $\bar t = \varepsilon_t \tilde{t}$.
The equation for the sphere surface can be approximated by $\varepsilon_{z_S} \tilde{z}_S = \frac{1}{2} \varepsilon^2\tilde{r}^2 - \varepsilon_t\tilde{t}$. 
As previously the determination of these scaling functions is obtained by dominant balance arguments: $\varepsilon_{z_S}= \varepsilon_t = \varepsilon^2$.
Note that at short times the intersection radius between the sphere and the impacting plane is given by $\tilde r_\mathrm{intersect} = \sqrt{2\tilde t}$.

\noindent We remark that as in the original study of Wagner, a scale separation between $\bar{z}_S$ and $\bar{z}$ exists \citep[\textit{small deadrise angle hypothesis}, see \textit{e.g.} ][]{Oliver2002}. This scale separation arises because the drop typical radius of curvature ($O(1)$) is very large in front of the other lengthscales of the problem, see Fig.~\ref{fig:scaling}.

\noindent We now turn on to the surface perturbation $\bar{h}(\bar{r},\bar{t})$, that embodies the impact-induced flow. Recalling that $\bar h$ represents a perturbation around a falling sphere, we can express the position of the free surface with the following implicit equation: $
 \mathcal{\bar S}(\bar{r},\bar{z},\bar{t})  =  \bar{z}-\left(\bar{z}_S-\bar{h}(\bar{r},\bar{t})\right)$.
Introducing an appropriate gauge function $\varepsilon_h$ such that $\bar{h} = \varepsilon_h \tilde{h}$ we obtain by dominant balance analysis that $\varepsilon_h = \varepsilon^2$. It follows that:
\begin{equation}
\begin{array}{rcl}
 \mathcal{ \tilde S}(\tilde{r},\tilde{z},\tilde{t}) & = & \displaystyle  \tilde{z} - \frac{1}{2} \tilde{r}^2 + \tilde{t} + \tilde{h}(\tilde{r},\tilde{t})\\[1em]
  & = & 0 \quad \text{on the free surface}.
\end{array}
\label{eq:free_surface_eq}
\end{equation}

\noindent The kinematic boundary condition derives from the previous equation. At the free surface, we have:
\begin{equation}
\frac{\mathrm d \mathcal {\bar S}}{\mathrm d \bar t} = 1  + \underbrace{\frac{\partial \tilde{h}}{\partial \tilde{t}}}_{\displaystyle\mathcal{O}(1)}  - \underbrace{ \displaystyle\varepsilon_{\phi} \tilde{r} \frac{\partial \tilde{\phi}}{\partial \tilde{r}}}_{\displaystyle\mathcal{O}(\varepsilon_{\phi})}  + \underbrace{ \varepsilon_{\phi} \frac{\partial \tilde{h}}{\partial \tilde{r}} \frac{\partial \tilde{\phi}}{\partial \tilde{r}}}_{\displaystyle\mathcal{O}(\varepsilon_{\phi})} + \underbrace{\frac{\varepsilon_{\phi}}{\varepsilon} \frac{\partial \tilde{\phi}}{\partial \tilde{z}}}_{\displaystyle\mathcal{O}(\varepsilon_{\phi}/\varepsilon)}  = 0,
\end{equation}
where $\bar \phi = \varepsilon_{\phi} \tilde{\phi}$. It is impossible here to keep all terms at the same order; the dominant balance between the vertical velocities $\partial \tilde{h}/\partial \tilde{t}$ and $\partial \tilde{\phi}/\partial \tilde{z}$ implies that $\varepsilon_{\phi} = \varepsilon$. At leading order, the kinematic boundary condition is therefore reduced to : 
\begin{equation}
1 + \frac{\partial \tilde{\phi}}{\partial \tilde{z}} + \frac{\partial \tilde{h}}{\partial \tilde{t}} = 0 \quad \text{on the free surface.}
\end{equation}
It proves convenient to introduce a translation of the velocity potential such that $\tilde{\phi} = -\tilde{z} + \check{\phi}$. This translation merely accounts to analyse the problem in the falling-drop reference frame. The kinematic boundary condition is then simply rewritten as:
\begin{equation}
\frac{\partial \tilde{h}}{\partial \tilde{t}} = - \frac{\partial \check{\phi}}{\partial \tilde{z}} \quad \text{on the free surface.}
\end{equation}
Inserting these different scaled variables into Bernoulli's equation, we obtain: \begin{equation}
\varepsilon_p \tilde p + \frac{1}{\varepsilon} \frac{\partial \check{\phi}}{\partial \tilde{t}} + \frac{1}{2} \left[ \left( \frac{\partial \check{\phi}}{\partial \tilde{r}} \right)^2 + \left(-1 + \frac{\partial \check{\phi}}{\partial \tilde{z}} \right)^2 \right]= \frac{1}{2} \quad \text{in the liquid},
\end{equation}
where $\bar p = \varepsilon_p \tilde p$.
The scale of the pressure, $\varepsilon_p = \frac{1}{\varepsilon}$, is here seen to be as large as the contact zone is small -- as expected in an impact problem.
At leading order, Bernoulli's equation is therefore reduced to:
\begin{equation}
\tilde p = - \frac{\partial \check{\phi}}{\partial \tilde{t}} \quad \text{in the liquid}.
\end{equation}

\begin{table}
\begin{center}
\begin{tabular}{@{}ccc@{}}
$\bar r  = \varepsilon \tilde r$ \qquad&
$\bar z  = \varepsilon \tilde z$ \qquad& 
$\bar t  = \varepsilon^2 \tilde t$\\[.1em]
$\bar p = \varepsilon^{-1} \tilde p$ \qquad&
$\bar \phi = \varepsilon \tilde \phi$ \qquad&  
$(\bar u, \bar v)  =  (\tilde u, \tilde v).$\\
\end{tabular}
\end{center}
\label{tab:tab0}
\caption{Résumé of the most important asymptotic scales of the problem.}
\end{table}

It follows from this equation that the constant pressure Dirichlet boundary condition on the free surface $p=0$ can be recast as a condition for the potential at the free surface: $\phi = \text{const}$, where the constant is arbitrary.
Without loss of generality, we set from now on this constant to zero. 

\noindent Finally, as classic in water wave theory, we exploit the shallowness of the gap between the free surface and the plane to transfer the boundary condition at the free surface onto the plane \citep[see \textit{e.g.} ][\S 3.8]{VanDyke1975}.

\begin{figure}
\centering
\includegraphics{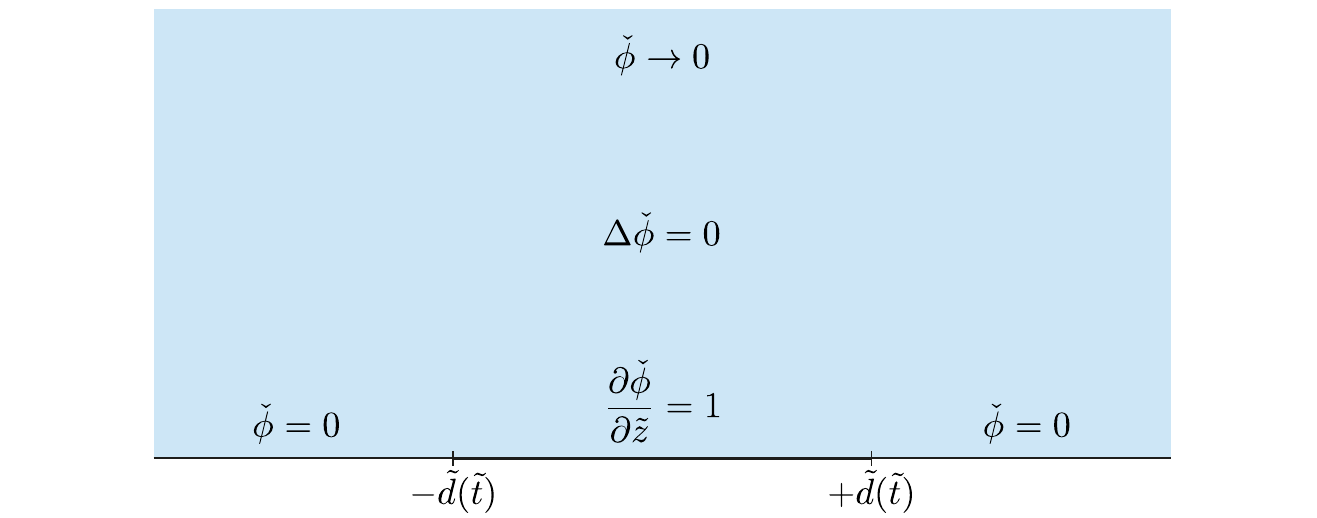}
\caption{Leading order outer problem for times of order $\varepsilon^2$.}
\label{fig:leading_order}
\end{figure}

\noindent Summarizing, the model problem takes the following expression:
\begin{alignat}{2}
 \frac{1}{\tilde{r}} \frac{\partial}{\partial \tilde{r}} \left( \tilde{r} \frac{\partial \check{\phi}}{\partial \tilde{r}} \right) + \frac{\partial^2 \check{\phi}}{\partial \tilde{z}^2} = 0 \quad & \text{in the liquid,}  \\
 - \frac{\partial \check{\phi}}{\partial \tilde{t}} = \tilde p \quad & \text{in the liquid,}\\
\intertext{the locus $\tilde d(\tilde t)$ of the contact line is determined with the Wagner condition:}
\tilde{h}(\tilde{r},\tilde{t}) = \frac{1}{2} \tilde{r}^2 -  \tilde{t} \quad & \text{for} \quad \tilde{r} = \tilde d(\tilde{t}),\\
\intertext{so that the boundary conditions at $\tilde z=0$ read:}
\check{\phi} = 0 \quad & \text{for} \quad \tilde{r} > \tilde d(t),\\
\frac{\partial \tilde{h}}{\partial \tilde{t}} = - \frac{\partial \check{\phi}}{\partial \tilde{z}} \quad & \text{for} \quad \tilde{r} > \tilde d(t), \\
\frac{\partial \check{\phi}}{\partial \tilde{z}} = 1  \quad  & \text{for} \quad \tilde{r} < \tilde d(\tilde{t}),\\
\intertext{and the far-field behaviour is given by:}
\check{\phi} \rightarrow 0 \quad & \text{as} \quad  \tilde{r}, \tilde{z} \rightarrow \infty \\
\tilde{h} \rightarrow 0 \quad & \text{as} \quad  \tilde{r}  \rightarrow \infty.
\end{alignat}
Finally the corresponding model geometry is sketched Fig.~\ref{fig:leading_order}.
\noindent We remark that the previous set of equations resembles to that of the classic water entry problem, and can be solved using the methodology described in \textit{e.g.} \citet{Oliver2002}. In the next section though we will present an alternate method based on self-similar solutions.

\section{Self-similar solutions and numerical simulations}
\label{sec:selfsimilarsolution}
\subsection{A self-similar problem}
\label{sec:selfsimilar}
To reveal the self-similar nature of our problem, we classically look in the following for scale invariance (\cite{darrofran82}). We start by expressing the fact that any variable $\tilde q$ in $(\tilde{r},\tilde{z},\tilde{t},\check{\phi}, \tilde{h},\tilde{d},\tilde p)$ can be rewritten as  $\tilde q = \lambda_q \hat q$, where $\hat q$ is a rescaled variable and $\lambda_q$ a numerical stretching coefficient embodying the change of scale.  
Inserting these variables into the governing equations, it is straightforward to see that invariance of Laplace equation through this stretching requires $\lambda_r = \lambda_z$.
Similarly, expressing the invariance of Wagner condition yields $\lambda_h = \lambda_t$,
$\lambda_r =\sqrt{\lambda_t}$ and $\lambda_d = \sqrt{\lambda_t}$. The same operation performed on the additional boundary conditions finally imposes $\lambda_\phi = \sqrt{\lambda_t}$ and $\lambda_p = 1/\sqrt{\lambda_t}$.  
Note that $\lambda_t$ remains here as the sole stretching parameter. 

\noindent The pressure field can be written as an implicit function of time and space as follows: $\mathcal{F}(\tilde p,\tilde{r},\tilde{z},\tilde{t}) = 0 $. Upon using the previous scale invariance arguments, this relation may be rewritten as $\mathcal{F}(\hat{p} / \sqrt{\lambda_t},\sqrt{\lambda_t} \hat{r},\sqrt{\lambda_t}\hat{z},\lambda_t \hat{t}) = 0 $.
A simple algebraic manipulation allows to remove the $\lambda_t$ dependence for all but one variables, so that finally $\mathcal{G}(\sqrt{\hat{t}} \; \hat{p},\hat{r}/\sqrt{\hat{t}},\hat{z}/\sqrt{\hat{t}},\lambda_t \hat{t}) = 0 $, for any $\lambda_t$. Remarking that for a given $\hat t$, this function has to cancel whatever the choice of the scale $\lambda_t$, it readily appears that the last variable is superfluous. In other words, a relation linking $\sqrt{\hat{t}} \; \hat{p}$ to $\hat{r}/\sqrt{\hat{t}}$ and $\hat{z}/\sqrt{\hat{t}}$ only must exist.

\noindent The pressure field may therefore be rewritten explicitly as:
\begin{equation}
\label{eq:psim}
\tilde{p} =
\frac{1}{\sqrt{\tilde{t}}}
\mathcal{P} \left(\frac{\tilde{r}}{\sqrt{\tilde{t}}}, \frac{\tilde{z}}{\sqrt{\tilde{t}}} \right).
\end{equation}
With a similar reasoning, and upon introducing the self-similar variables $\xi=\tilde{r}/\sqrt{\tilde{t}}$ and $\eta = \tilde{z}/\sqrt{\tilde{t}}$, 
 we readily obtain :
\begin{equation}
\check{\phi}(\tilde{r},\tilde{z},\tilde{t}) = \sqrt{\tilde{t}} \; \Phi(\xi,\eta), \quad \tilde{h}(\tilde{r},\tilde{t}) = \tilde{t} \; \mathcal{H}(\xi) \quad \text{and} \quad
\tilde{d}(\tilde x,\tilde t) =  \sqrt{\tilde t}\; \delta,
\end{equation}
where $\Phi$ and $\mathcal{H}$ are unknown functions of the self-similar variables and $\delta$ a constant representing the (fixed) position of the contact line in self-similar space. 
This allows us to formulate the self-similar version of the drop impact problem : 
\begin{alignat}{2}
 \frac{1}{\xi} \frac{\partial}{\partial \xi} \left( \xi \frac{\partial \Phi}{\partial \xi} \right) + \frac{\partial^2 \Phi}{\partial \eta^2} = 0 \quad & \text{in the liquid,}  \\
\mathcal{P}(\xi,\eta) = \frac{1}{2} \left( - \Phi(\xi,\eta) + \xi \frac{\partial \Phi}{\partial \xi} + \eta \frac{\partial \Phi}{\partial \eta} \right) \quad & \text{in the liquid,} \label{eq:selfsimilarpressure}\\
\intertext{the boundary conditions at $\eta = 0$ take the following form:}
\mathcal{H}-\frac{1}{2} \xi \frac{\partial \mathcal{H}}{\partial \xi} = - \frac{\partial \Phi}{\partial \eta} \quad & \text{for} \quad \xi > \delta,\label{eq:BCkinematicselfsimilar} \\
\frac{\partial \Phi}{\partial \eta} = 1   \quad & \text{for} \quad \xi < \delta,\label{eq:BCwallselfsimilar}\\
\Phi = 0 \quad & \text{for} \quad \xi > \delta,\label{eq:BCfreesurfaceselfsimilar}\\
\intertext{the far-field behaviour is:}
\Phi \rightarrow 0 \quad & \text{as} \quad \xi, \eta \rightarrow \infty \\
\mathcal{H} \rightarrow 0 \quad & \text{as} \quad  \xi \rightarrow \infty,
\label{eq:farfieldHselfsimilar}
\intertext{and the self-similar version of Wagner condition is finally given by:}
\mathcal{H}(\xi) = \frac{1}{2} \xi^2 - 1 \quad & \text{for} \quad \xi = \delta.
\end{alignat}
This problem can now be solved in several steps.

\subsection{Self-similar potential}
\label{sec:potential}
In this geometry, Laplace equation can be solved with variable separation, leading to a family of elementary cylindrical harmonic solutions with an exponential behaviour in $\eta$ and an oscillatory one in $\xi$. We recompose by summation and obtain :
\begin{equation}
\Phi(\xi,\eta) = \int_0^{\infty} \!\mathcal{C}(k)  J_0 (k \xi) e^{-k \eta}\,\mathrm{d}k.
\end{equation}
The weight function $\mathcal{C}(k)$ is determined with boundary conditions~(\ref{eq:BCwallselfsimilar}) and~(\ref{eq:BCfreesurfaceselfsimilar}), leading to the following pair of dual integral equations:
\begin{subnumcases}{}
\int_0^{\infty} k \mathcal{C}(k) J_0(k \xi) \,\mathrm dk = -1 & for 
$\xi < \delta$, \\
\int_0^{\infty} \mathcal{C}(k) J_0(k \xi) \,\mathrm dk = 0 & for $\xi > \delta$.
\end{subnumcases}
Solving these dual integral equations using the technique described in \citet{Sneddon1960}, we obtain a closed-form expression for the weight function:
\begin{equation}
\mathcal{C}(k) = \frac{2}{\pi}
\frac{\delta k \cos(k \delta)-\sin(k \delta)}{k^2}=
\frac{2}{\pi}
\frac{\mathrm d}{\mathrm dk} \left(\frac{\sin(k \delta)}{k}\right).
\end{equation}
Anticipating the description of the contact line dynamics, we now derive $\partial \Phi/\partial \eta$ at the substrate level $\eta=0$:
\begin{equation}
\frac{\partial \Phi}{\partial \eta}
=
 - \frac{2}{\pi}\int_0^{\infty}  
  \frac{k \delta \cos(k \delta)-\sin(k \delta)}{k^2}
  J_0 (k \xi) k \, \mathrm dk,
\end{equation}
where we recognize the sum of two Hankel transforms \citep[see \textit{e.g.}][table IV, page 528]{Sneddon1951}. This allows us to obtain the following explicit expression for $\partial \Phi/\partial \eta$ for $\eta = 0$:
\begin{equation}
\frac{\partial \Phi}{\partial \eta}
=
1\;
 \text{for $\xi<\delta$\quad and}\quad
\frac{\partial \Phi}{\partial \eta}
=
-\frac{2}{\pi}
\left(
\frac{\delta}{\sqrt{\xi^2-\delta^2}} - \arcsin\left(\frac{\delta}{\xi}\right) \right)  \;\text{for $\xi>\delta$.}
\label{eq:verticalvelocityalongsubstrate}
\end{equation}

\subsection{Wagner condition and contact line dynamics}
With the help of the vertical velocity expression in the near wall region just derived, we can rewrite the kinematic boundary condition~(\ref{eq:BCkinematicselfsimilar}) as:
\begin{equation}
\mathcal{H}(\xi)-\frac{1}{2} \xi \frac{\partial \mathcal{H}}{\partial \xi}(\xi) =  \frac{2}{\pi}
\left(\frac{\delta}{\sqrt{\xi^2-\delta^2}} - \arcsin\left(\frac{\delta}{\xi}\right) \right)  \;\text{for $\xi>\delta$.}
\end{equation}
This inhomogeneous differential equation can be solved using variation of parameters, \textit{i.e.} looking for a solution of the form $\mathcal H(\xi)=\xi^2 f(\xi)$. This gives:
\begin{equation}
\bigg[f(\xi)\bigg]_\delta^{+\infty}
=-\frac{2}{\pi}\int_{\delta}^{\infty} \frac{2}{ \xi^3} \left(
\frac{\delta}{\sqrt{\xi^2-\delta^2}} - \arcsin\left(\frac{\delta}{\xi}\right) \right)  \mathrm d\xi.
\end{equation}
Upon using the far-field decaying behaviour of $\mathcal H$ (see equation~(\ref{eq:farfieldHselfsimilar})), this last equation reduces to $f(\delta) =\frac{1}{2} \delta^{-2}$ so that at the contact line the drop deformation is:
\begin{equation}
\mathcal H(\delta)=\delta^2 f(\delta) = \frac{1}{2}.
\end{equation}
In the self-similar space, the Wagner condition therefore takes the following remarkably simple form:
\begin{equation}
\frac{1}{2}=\frac{1}{2} \delta^2 - 1,
\end{equation}
from which we finally derive the position of the contact line: 
\begin{equation}
\delta = \sqrt{3}.
\end{equation}

\begin{figure}
\centering
\includegraphics{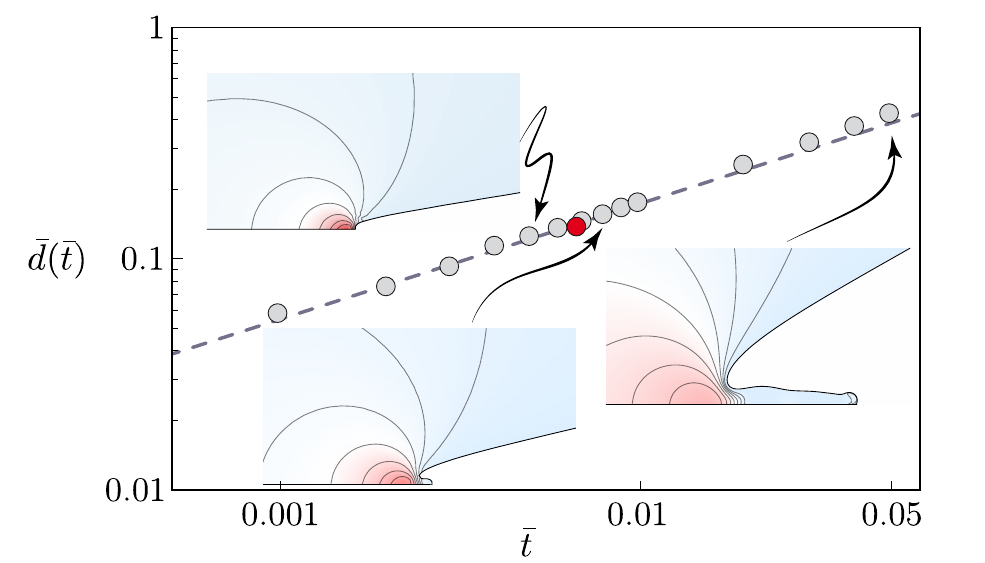}
\caption{Comparison between the theoretical position of the contact line as a function of time deduced from Wagner theory 
$\tilde d(\tilde t)=\sqrt{3 \tilde t}$ (dashed line) and the position of the contact line extracted from \gerris\ computations of an impacting drop at $\text{Re}=5000$ and $\text{We}=250$  (grey dots). The red dot marks the birth of the corolla.}
\label{fig:dt}
\end{figure}

It is interesting to remark that the contact line motion $\tilde d(\tilde t) = \sqrt{3 \tilde t}$ just predicted within the framework of Wagner theory is quite close from the rough truncated sphere approximation $\tilde r_\text{intersect} = \sqrt{2\tilde t}$ \citep{Rioboo2002}. Furthermore, and as for other liquid impact problems, we note that the agreement between this prediction and observations is fairly satisfactory. Indeed Fig.~\ref{fig:dt} reports early post-impact successive positions of the contact line extracted from numerical simulations performed with \gerris\ along with our theoretical prediction. Noticeably the superposition between theory and numerical results is excellent, at least until the moment of formation of a liquid corolla (here indicated with a red dot). 


\subsection{Analogy with the normal motion of an expanding disk in an infinite mass of liquid}
\label{sec:expanding_disk}
In \S \ref{sec:analogywaterentry} we proposed to visualize the flow in an impacting drop as the one induced by a flat rising disk expanding radially as the wet area (see also Fig.~\ref{fig:expanding_plate}). We are now in a position to formally justify this water-entry analogy. The axisymmetric flow induced by \textsl{`the motion of a thin circular disk with velocity $U$ normal to its plane, in a infinite mass of liquid'} is for example analysed in Lamb's classic textbook~\S 101 \citep{Lamb1975}.
After deriving some elementary axisymmetric solutions of Laplace equation of the form $\exp(\pm k z) J_0(k r)$ in \S 100, \citeauthor{Lamb1975} examined a variety of axisymmetric potential flows. Among those was the one (later connected to the flow around a flat circular disk in normal incidence) where at the symmetry plane $z=0$ the potential takes the value $\phi=C \sqrt{a^2-r^2}$ for $r<a$ and $\phi=0$ for $r>a$, with $a$ the disk radius. The solution for this problem was stated under the following integral representation:
\begin{equation}
\phi(r,z) = -C \int_0^{\infty} \text{e}^{-k z} J_0(k r) \frac{\mathrm d}{\mathrm dk}\left(\frac{\sin k a}{k}\right) \mathrm dk.
\end{equation}
And from \textsl{`a known theorem in Electrostatics'}, Lamb obtained the expression for the vertical velocity in the symmetry plane: 
\begin{subnumcases}{-\left(\frac{\partial \phi}{\partial z}\right)_{z=0} = }
\frac{1}{2} \pi C & for $r<a$,\\
C \left(\arcsin\left(\frac{a}{r}\right)-\frac{a}{\sqrt{r^2-a^2}}\right)  & for $r>a$.
\end{subnumcases}
This corresponds precisely to the flow within the impacting drop, after posing $C = -2/\pi$ and $a=\delta$, thereby justifying formally our initial analogy between the impact-induced flow with the one associated with a flat rising disk rapidly expanding with the wet area. 
Setting $C = 2 U /\pi$, Lamb remarked that the above potential indeed describes the flow winding around a flat disk moving at velocity $U$. He further noted that a simple expression for the fluid half-space kinetic energy could be derived from the previous relation: 
\begin{equation}
T_\text{disk} = \frac{4}{3} \rho a^3 U^2.
\end{equation}
This expression can immediately be transposed into the (nondimensional) kinetic energy of the impact-induced flow within the drop:
\begin{equation}
\tilde T = 4\sqrt{3} \tilde t^{3/2},
\end{equation}
or, equivalently, into its dimensioned counterpart:
\begin{equation}
T = 4\sqrt{3} \rho U^{7/2} R^{3/2} t^{3/2}.
\label{eq:kineticenergy}
\end{equation}
We emphasize that this expression is derived within the frame of the falling drop and, as such, represents the kinetic energy of the defect flow associated with impact. Although a direct physical interpretation of this quantity is not straightforward, we will see in \S \ref{sec:normalforce} that the knowledge of this defect kinetic energy will allow for a direct determination of the impacting force.

\subsection{Structure of the velocity field}
\label{sec:velocity}
We now investigate the structure of the velocity field in the contact region and search for exact closed-form expressions and convenient approximations for this field.
\subsubsection{Integral representation of the velocity field}
\noindent In the fixed frame, the velocity field $\tilde{\boldsymbol{u}}(\tilde r,\tilde z,\tilde t)$ inside the impacting drop can formally be derived from the (untranslated) potential $-\eta +\Phi$. 
Following the arguments developed in \S \ref{sec:selfsimilar}, this velocity field is simply related to the self-similar velocity field $\boldsymbol{\mathcal{U}}(\xi,\eta)$ via the relation: 
\begin{equation}
\tilde{\boldsymbol{u}}(\tilde r,\tilde z,\tilde t) = \boldsymbol{\mathcal{U}}(\xi,\eta).
\end{equation}
where the components of the self-similar vector field $\boldsymbol{\mathcal{U}}=(\mathcal U_\xi,\mathcal U_\eta)$ are:
\begin{subnumcases}{}
\mathcal{U}_{\xi}(\xi,\eta)  =  \frac{\partial \Phi}{\partial \xi}, & \\
\mathcal{U}_{\eta}(\xi,\eta) = -1 + \frac{\partial \Phi}{\partial \eta}. &
\end{subnumcases}
Inserting the expression of the self-similar potential determined previously yields the following integral representation for the vector field components:
\begin{subnumcases}{\label{eq:integralrepresentationselfsimilarvelocity}}
\mathcal{U}_{\xi}(\xi,\eta)  =  - \frac{2}{\pi} \int_0^{\infty} \frac{\sqrt{3} k \cos(\sqrt{3} k)-\sin(\sqrt{3} k)}{k} e^{-k \eta} J_1 (k \xi) \, \mathrm dk, & \label{eq:radialvelocityinselfsimilarspace} \\
\mathcal{U}_{\eta}(\xi,\eta) = -1 - \frac{2}{\pi} \int_0^{\infty} \frac{\sqrt{3} k \cos(\sqrt{3} k)-\sin(\sqrt{3} k)}{k} e^{-k \eta} J_0 (k \xi) \, \mathrm dk. &
\end{subnumcases}
A closed-form expression is unfortunately not accessible in the general case. In the following however we calculate the value of these integrals at some particular places. 
\subsubsection{Closed-formed expressions for the velocity field along the axis and the substrate}
\label{sec:closedformvelocity}
\begin{figure}
\centering
\includegraphics{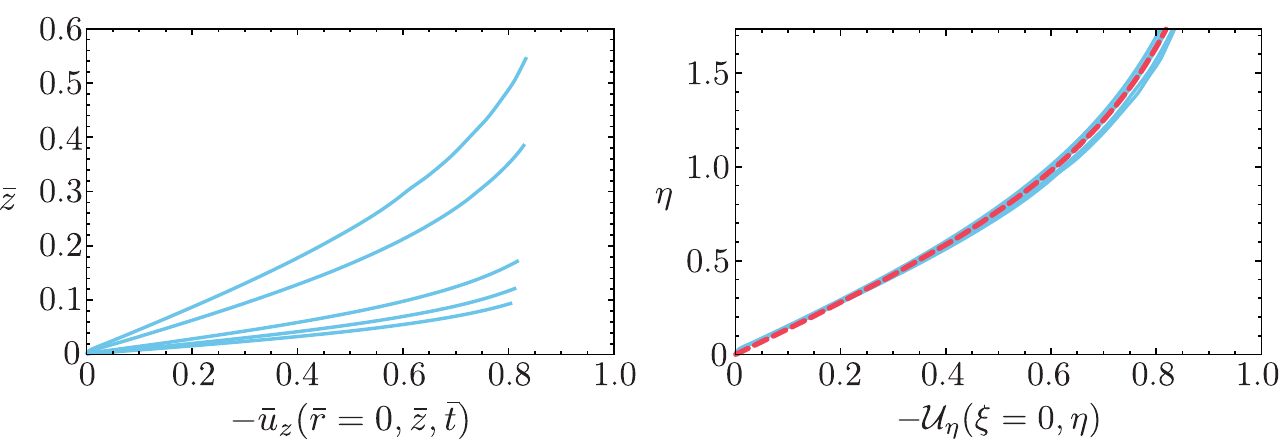}
\caption{Left : Axial velocity profiles along the axis extracted from \gerris\ computations at times $\bar{t} = 3\times 10^{-3}, 5\times 10^{-3}, 10^{-2}, 5\times 10^{-2}$ and $10^{-1}$. Right : Comparison between the analytical prediction for the axial velocity given by equation (\ref{eq:axialvelocityprofileselfsimilar}) (red dashed line) and numerical solutions obtained with \gerris , rescaled in the self-similar space (blue solid lines). The physical parameters for this simulation are $\text{Re}=5000$ and $\text{We}=250$.}
\label{fig:AxialVelocity}
\end{figure}

\noindent Simple analytical solutions for the velocity field can be obtained from~(\ref{eq:integralrepresentationselfsimilarvelocity}) at precise locations. Along the symmetry axis for example, where $\xi = 0$, the properties of integrals of exponentials allow to write:
\begin{equation}
\mathcal{U}_{\eta}(\xi=0,\eta) = -1 + \frac{2}{\pi} \left(\arctan \left(\frac{\sqrt{3}}{\eta} \right) - \frac{\sqrt{3} \eta}{3 + \eta^2} \right) \quad \text{ for } \eta \geq 0.
\label{eq:axialvelocityprofileselfsimilar}
\end{equation}
This last result is confronted with numerical velocity profiles extracted from \gerris\ computations in Fig.~\ref{fig:AxialVelocity}. The nice agreement between the theoretical solution and the numerical profiles seen in the self-similar space (Fig.~\ref{fig:AxialVelocity}b) here holds over more than a decade in time.

\noindent Analogously, analytical forms for~(\ref{eq:integralrepresentationselfsimilarvelocity}) can also be obtained along the substrate plane $\eta = 0$ by exploiting the properties of Hankel transforms \citep{Sneddon1951}. An expression for the vertical velocity is already provided with equation~(\ref{eq:verticalvelocityalongsubstrate}), after inserting $\delta = \sqrt{3}$. We remark that this analytical solution elucidates the faster-than-free-flight motion of the free surface near the contact line discerned in \S \ref{sec:contactlinenumericalobservations}. Likewise the radial velocity distribution across the wet area is found to be:
\begin{equation}
\mathcal{U}_{\xi}(\xi,\eta=0) = \frac{2}{\pi}\frac{\xi}{\sqrt{3 - \xi^2}} \quad \text{ for } 0 \leq \xi < \sqrt{3}.
\label{eq:slipvelocity}
\end{equation}
This unphysical inviscid slip velocity $\tilde{u}_e(\xi)$ cannot be observed in our simulations encompassing viscous effects. But this quantity is nonetheless relevant for it corresponds to the edge velocity of the viscous boundary layer (studied in detail in \S \ref{sec:viscous}).

\subsubsection{An unusual stagnation point flow}
\noindent In the very vicinity of the origin, the first order power series of the velocity field~(\ref{eq:integralrepresentationselfsimilarvelocity}) reads:
\begin{subnumcases}{}
\mathcal{U}_{\xi}(\xi,\eta)  \simeq \frac{2}{\pi \sqrt{3}} \xi, & \\
\mathcal{U}_{\eta}(\xi,\eta) \simeq -\frac{4}{\pi \sqrt{3}} \eta, &
\end{subnumcases}
or, equivalently, in dimensioned variables:
\begin{subnumcases}{}
u_{r}(r,z,t) \simeq \frac{2}{\pi \sqrt{3}} \; \sqrt{\frac{U}{R}} \frac{r}{\sqrt{t}}, & \\
u_{z}(r,z,t) \simeq -\frac{4}{\pi \sqrt{3}} \; \sqrt{\frac{U}{R}} \frac{z}{\sqrt{t}}. &
\end{subnumcases}
Though simple, this peculiar structure for the impact-induced unsteady stagnation point flow is nonetheless counter-intuitive and could not have been inferred from simple dimensional analysis. Noteworthy enough, this result is at variance with the typical structure of the later intermediate flow associated with spreading  $\bar{\boldsymbol u} \simeq (\bar r/\bar t,-2 \bar z/\bar t)$ \citep[see \textit{e.g.}][]{Eggers2010a,Lagubeau2012,Yarin1995}.  
\subsubsection{Beyond the stagnation point: a remark on the overall velocity field structure}
\begin{figure}
\centering
\includegraphics[]{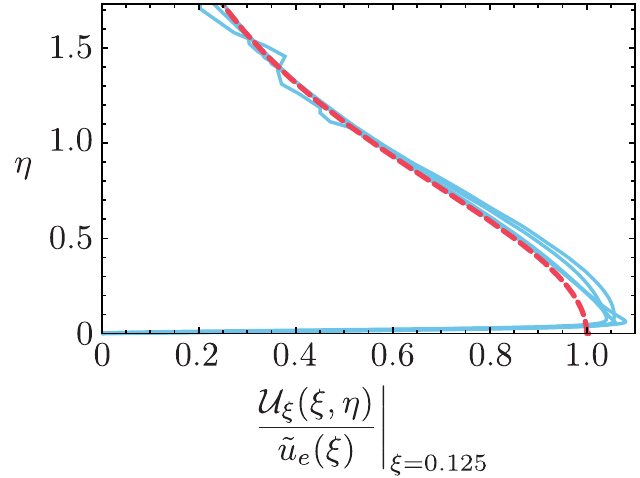} 
\includegraphics[]{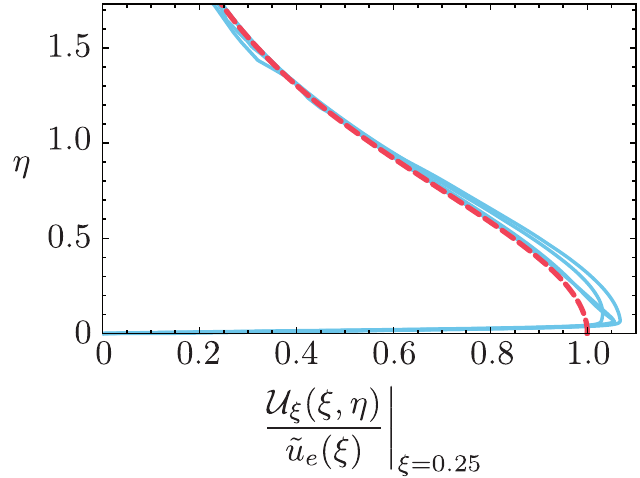}
\includegraphics[]{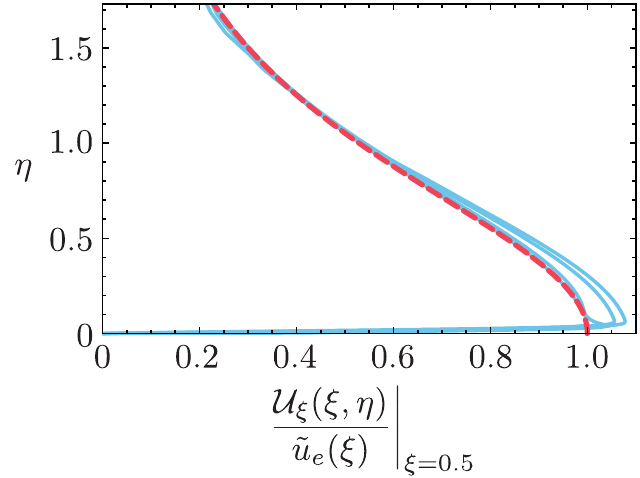}
\includegraphics[]{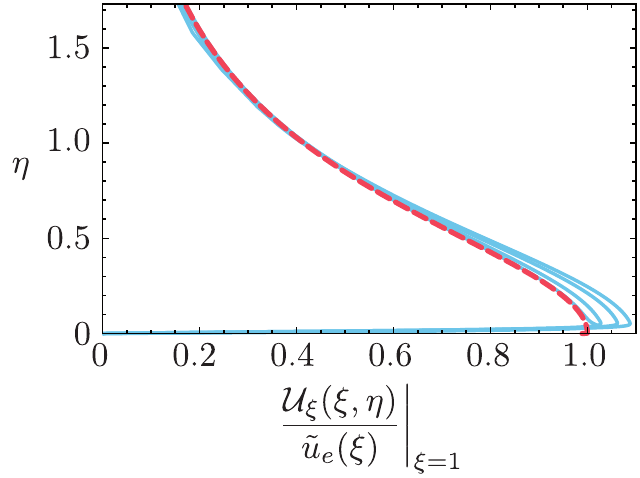}
\includegraphics[]{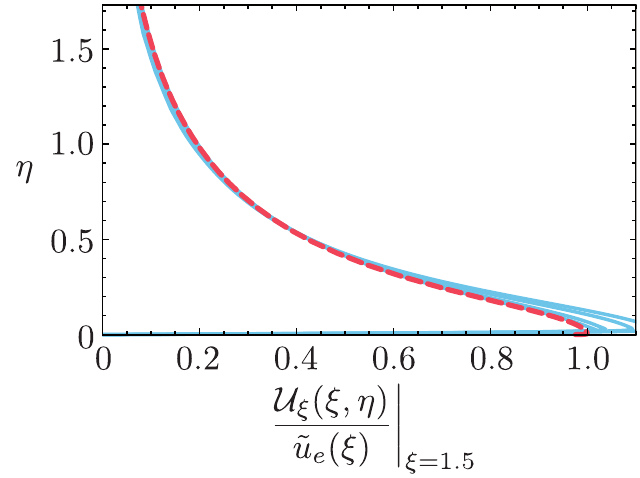}
\caption{Self-similar radial velocity as a function of $\eta$ for $\xi =$ 0.125, 0.25, 0.5, 1 and 1.5. These velocities are rescaled by the outer solution of the boundary layer $\tilde{u}_e(\xi)$ given by equation~(\ref{eq:slipvelocity}). Blue solid lines represent the numerical solutions extracted from \gerris\ computations in the self-similar space for $\bar{t} = 5\times 10^{-3}, 10^{-2}, 5\times 10^{-2}$ and $10^{-1}$ ($\text{Re}=5000$ and $\text{We}=250$). The red dashed line represents the theoretical solution $\mathcal{U}_{\xi}(\xi,\eta)$ given by equation~(\ref{eq:radialvelocityinselfsimilarspace}). Note that the boundary layer is so thin that it is almost indistinguishable (see also Fig.~\ref{fig:CoucheLimite}).}
\label{fig:VitesseRadiale}
\end{figure}

\begin{figure}
\centering
\includegraphics{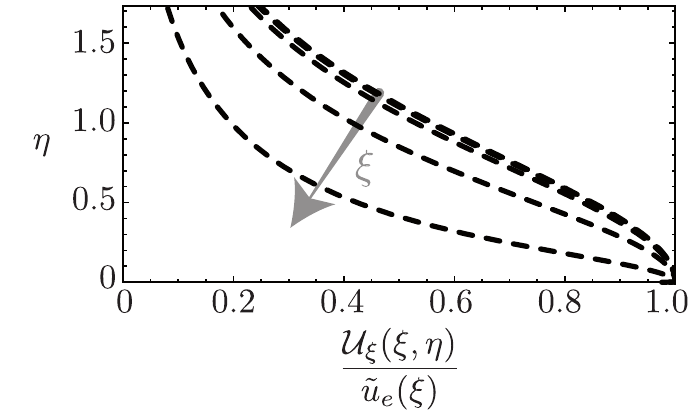}
\caption{Evolution of the analytical self-similar radial velocity 
given by equation~(\ref{eq:radialvelocityinselfsimilarspace})
as a function of $\eta$ for $\xi =$ 0.125, 0.25, 0.5, 1 and 1.5. These velocities are rescaled by the outer solution of the boundary layer $\mathcal{U}_{\xi}(\xi,\eta=0) = \tilde{u}_e(\xi)$, equation (\ref{eq:slipvelocity}). 
}
\label{fig:XiVariable}
\end{figure}

\noindent The previous approximation for the impact flow is valid in a small region near the origin. To further investigate the limits of this representation we show Fig.~\ref{fig:VitesseRadiale} different radial velocity profiles corresponding to various locations $\xi$. The collapse of the numerical profiles taken at various $\bar r$ and $\bar t$ (but such that $\bar r / \sqrt{\bar t} = \tilde r / \sqrt{\tilde t}$ is constant in each figure) onto the theoretical profiles is again an illustration of the relevance of the self-similar representation. But it is also to be noted that while the stagnation point ansatz disregards any radial velocity variation in $\eta$, the profiles exhibit a sensible variation along the vertical coordinate $\eta$. 
This variation is best depicted with Fig.~\ref{fig:XiVariable} where theoretical radial velocity profiles taken at different values $\xi$ have been represented. Noteworthy enough, profiles corresponding to $\xi \lesssim  1$ collapse on a single curve. We therefore speculate the following functional dependence for the velocity $\mathcal{U}_{\xi}(\xi,\eta) = u_e(\xi) f(\eta)$ in this region. For larger values of $\xi$ though, significant deviations from this behaviour arise and variable separation cease to hold:  $\mathcal{U}_{\xi}(\xi,\eta) = \tilde u_e(\xi) g(\xi,\eta)$ with $g(\xi,\eta=0)=1$ for $\xi \gtrsim 1$.

\subsubsection{Flow pattern, contact line bypass and Lamb analogy}
\noindent We now define the self-similar stream function $\Psi(\xi,\eta)$ in the drop reference frame from the potential:
\begin{subnumcases}{}
\frac{1}{\xi} \frac{\partial \Psi}{\partial \eta} = \frac{\partial \Phi}{\partial \xi} , & \\
-\frac{1}{\xi} \frac{\partial \Psi}{\partial \xi} = \frac{\partial \Phi}{\partial \eta}.&
\end{subnumcases}
By integration of the previous relations we deduce the following expression for $\Psi(\xi,\eta)$:
\begin{equation}
\Psi(\xi,\eta) = \frac{2}{\pi} \int_0^{\infty} \frac{\sqrt{3} k \cos(\sqrt{3} k)-\sin(\sqrt{3} k)}{k^2} \text{e}^{-k \eta} \xi J_1 (k \xi) \;\mathrm dk,
\label{eq:psiFF}
\end{equation}
up to a constant.
\begin{figure}
\centering
\includegraphics[width=10cm]{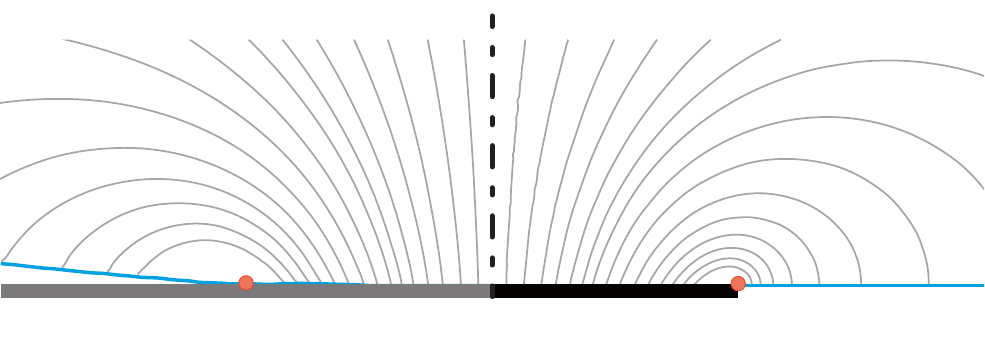}
\caption{Comparison between the flow pattern within an impacting drop (left) and around a rapidly expanding disk (Lamb analogy, right) in the self-similar space. In both cases, the streamlines are represented in the moving frame. The red dots represent the theoretical position of the contact line $\xi = \sqrt{3}$. The numerical streamlines represented on the left are derived from the velocity field computed with \gerris\ at $\bar{t}=10^{-3}$ ($\text{Re}=5000$ and $\text{We}=250$). The theoretical streamlines shown on the right correspond to isovalues of $\Psi(\xi,\eta)$ defined in equation~(\ref{eq:psiFF}) (note the correspondence with Lamb's figure page 145).   }
\label{fig:Streamlines}
\end{figure}
Formally, $\Psi(\xi,\eta)$ is the stream function describing the winding flow around a flat rising disk \citep[][\S 108]{Lamb1975}. Figure~\ref{fig:Streamlines} offers a comparison between the streamlines of this Lamb analogy and the ones computed with \gerris\ for the drop impact problem in the self-similar space. A good qualitative agreement between the analytical and the numerical streamlines is noticeable, comforting the expanding disk analogy followed here. Interestingly the winding motion around the contact line, as well as the falling velocity overshoot near this region, are both captured with this analogy and can be correlated with the peculiarities of the winding flow near the edge of a rising disk.

\subsection{Self-similar pressure }
\label{sec:pressure}
From the knowledge of the velocity potential we are now in a position to derive the pressure field as the time derivative of the  potential. In the self-similar space, the pressure field is given by equation (\ref{eq:selfsimilarpressure}).
\begin{figure}
\centering
\includegraphics{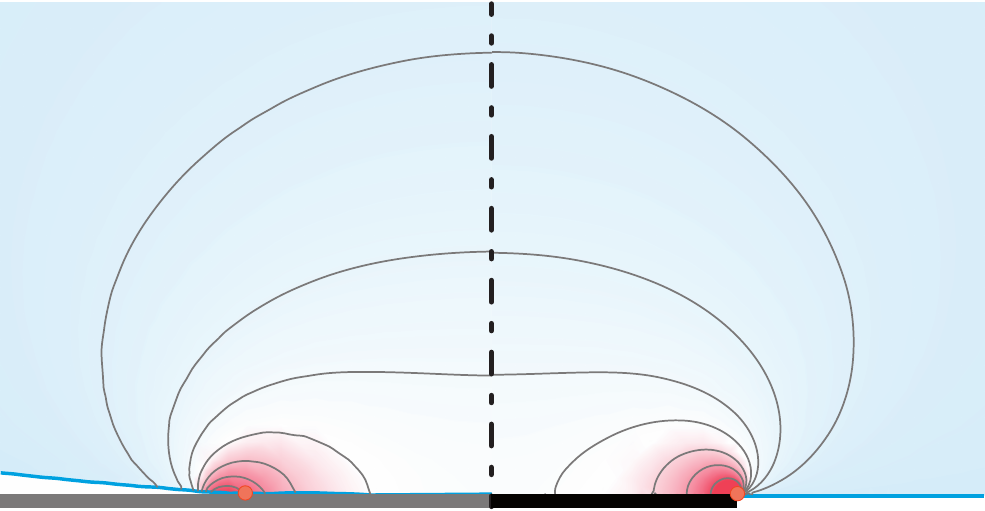}
\caption{Comparison between the pressure field developing inside an impacting drop (left) and around a rapidly expanding disk (Lamb analogy, right). The pressure field represented on the left is extracted from \gerris\ computations and represented in the self-similar space ($\bar{t}=10^{-3}$, $\text{Re}=5000$, $\text{We}=250$; The isovalues are: 0.12, 0.24, 0.36, 0.48, 0.6, 0.72, 0.84). The self-similar theoretical pressure field represented on the right is given by equation (\ref{eq:selfsimilarpressure}) (isovalues: 0.13, 0.28, 0.445, 0.57, 0.73, 0.9, 1.2).
Though isovalues have been slightly changed between the two panels, theoretical and numerical results are in a good overall agreement.
}
\label{fig:IsoPression}
\end{figure}
Figure~\ref{fig:IsoPression} proposes a comparison between the structure of the self-similar pressure extracted from numerical computations performed with \gerris\ and the theoretical prediction. There it can be seen that the overall structure of the pressure field developing in the impacting drop, and in particular the pressure peak in the vicinity of the contact line already pinpointed out in Fig.~\ref{fig:tryptic}, nicely matches with the theory.   
Interestingly, the structure just described is at variance with the pressure distribution around a flat disk rising steadily (Lamb's original problem). Indeed in such a configuration the pressure is expected to be maximal in the stagnation point area, whereas in our model problem the pressure peaks near the contact line/disk edge. This is a consequence of the motion unsteadiness: the pressure is here dominated by the $\partial \check \phi / \partial \tilde t$ contribution rather than the steady $\frac{1}{2}\tilde \nabla \check \phi^2$ term. 

\begin{figure}
\centering
\includegraphics[width=6cm]{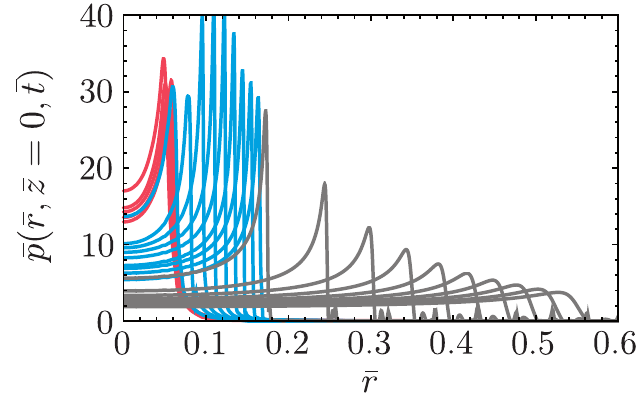}
\includegraphics[width=6cm]{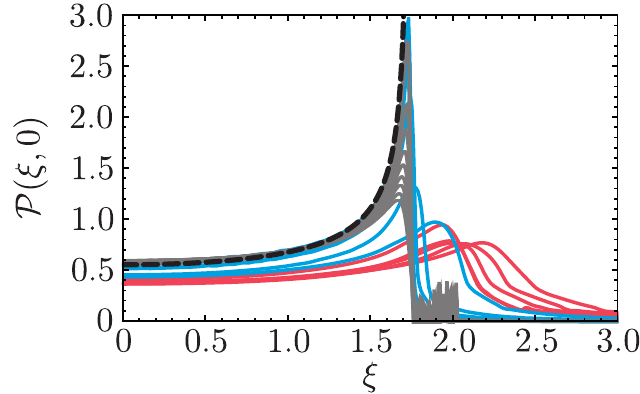}
\caption{Left: Pressure trace on the substrate $\bar z=0$ obtained from the numerical simulations between $\bar{t} = 5\times 10^{-4}$ and $10^{-1}$ for $\text{Re}=5000$ and $\text{We}=250$. The color code for each decade is the same as in Fig.~\ref{fig:p0}. Note that the curves are equally distributed within each decade. Right: Same as in the left but in the self-similar space. The black dashed curve represent the analytical solution for $\mathcal{P}(\xi,\eta=0)$.}
\label{fig:RadialPressure}
\end{figure}

As in \S \ref{sec:closedformvelocity}, closed form expressions for the pressure can be obtained along the axis and the substrate plane. The radial structure of the self-similar pressure across the wet area reads $\mathcal{P}(\xi,\eta=0) = \frac{3}{\pi \sqrt{3-\xi^2}}$ for $0 \leq \xi<\sqrt{3}$. This analytical prediction is confronted with \gerris\ numerical results in Fig.~\ref{fig:RadialPressure}. 
After a transient numerical initialization phase (corresponding to the red curves), the pressure profiles collapse on the self-similar analytical solutions (blue curves). In accordance with the overall pressure field structure depicted earlier, the pressure radial profile presents a local minimum at $\xi = 0$ and a maximum in the vicinity of the contact line, that is for $\xi = \sqrt{3}$ -- where the analytical solution exhibits an inverse square-root singularity. We note that for later times the pressure peak is smoothed out in the numerical simulations (grey curves). As this regularization coincides with the birth of the ejecta sheet, we conjecture that this fall-off can appropriately be described with a second-order Wagner theory \citep{Korobkin2007,Oliver2007}.

\begin{figure}
\centering
\includegraphics[width=6cm]{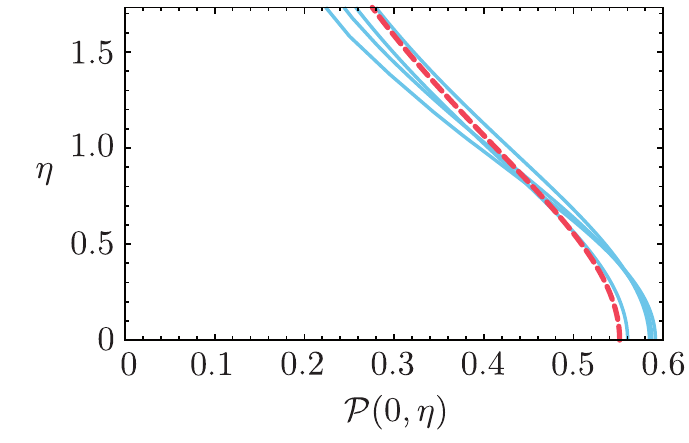}
\caption{Pressure along the axis $\bar r = 0$ obtained with \gerris\ for $\bar{t} = 5\times 10^{-3}, 10^{-2}, 5\times 10^{-2}$ and $10^{-1}$ (same as in Fig.~\ref{fig:AxialVelocity} right) represented in the self-similar space ($\text{Re}=5000$ and $\text{We}=250$). The red dashed line is the analytical solution for $\mathcal{P}(\xi=0,\eta)$. Fluctuations of the pressure around the theoretical prediction is to be related with numerical projections errors.}
\label{fig:PressionAxiale}
\end{figure}

Similarly the expression for the self-similar pressure along the symmetry axis can also be obtained analytically:  $\mathcal{P}(\xi=0,\eta) = \frac{3 \sqrt{3}}{\pi (3+\eta^2)}$ for $\eta \geq 0$. Figure~\ref{fig:PressionAxiale} compares this last result with rescaled axial pressure profiles extracted from numerical simulations. There again the agreement between the computations and the theory is seen to hold for a large time span.
 
The structure of the pressure field in the vicinity of the origin can be inferred from these last results. Reexpressing the pressure cuts determined in terms of $\tilde{r}$, $\tilde{z}$ and $\tilde{t}$, we get:
\begin{subnumcases}{}
\tilde{p}(\tilde{r},\tilde{z}=0,\tilde{t}) = \frac{3}{\pi \sqrt{3 \tilde{t} - \tilde{r}^2}}, & \\
\tilde{p}(\tilde{r}=0,\tilde{z},\tilde{t}) =  \frac{3 \sqrt{3 \tilde{t}}}{\pi (3 \tilde{t} + \tilde{z}^2)}.&
\end{subnumcases}
From these two relations, we deduce the following expansion for the pressure near the origin:
\begin{equation} 
\tilde{p}(\tilde r,\tilde z,\tilde{t}) = \frac{\sqrt{3}}{\pi} \; \tilde{t}^{-\frac{1}{2}}
\left(1 + \frac{\tilde r^2}{6 \tilde t}- \frac{\tilde z^2}{3 \tilde t}\right)+ \dots
\end{equation} 
This expression provides with a local approximation for 
$\partial \check \phi/\partial \tilde t$ from which, after time integration and space differentiation, we readily recover the stagnation point flow structure found earlier: $(\tilde u_{\tilde r},\tilde u_{\tilde z}) =
\frac{2}{\pi \sqrt{3}} (\tilde r/\sqrt{\tilde t},
-2 \tilde z/\sqrt{\tilde t}).$
This near-axis behaviour emphasizes again that simple intuitive dimensional analysis suggestion 
$r/t$ and $-z/t$ is here not relevant.  

\begin{figure}
\centering
\includegraphics[width=6cm]{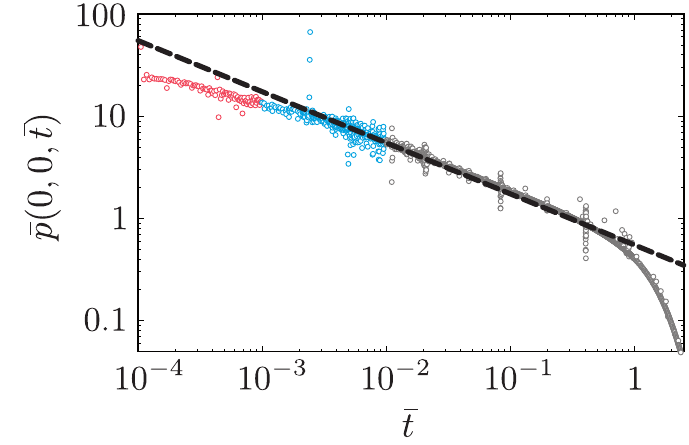}
\caption{Time evolution of the pressure $\bar{p}(0,0,\bar{t})$ measured at the origin in the numerical simulations with \gerris\ ($\text{Re}=5000$ and $\text{We}=250$). Note that each decade is represented with a different colour. The theoretical prediction $\bar{p}(0,0,\bar{t}) = \frac{\sqrt{3}}{\pi} \; \bar{t}^{-\frac{1}{2}}$ is superimposed with a black dashed line.}
\label{fig:p0}
\end{figure}

The leading order term for the pressure at the origin follows: 
\begin{equation}
\tilde{p}(0,0,\tilde{t}) = \frac{\sqrt{3}}{\pi} \; \tilde{t}^{-\frac{1}{2}},
\text{ or, with dimensions }
p(0,0,t) =\frac{\rho U^{3/2}}{\pi} 
\label{eq:PressionOrigine}
\sqrt{\frac{3 R}{t}}.
\end{equation} 
This result extends the $t^{-\frac{1}{2}}$ scaling law proposed by \citet{Josserand2003} on the basis of scaling arguments.
A comparison between this theoretical prediction and \gerris\ numerical simulations is proposed Fig.~\ref{fig:p0}, using the color code of Fig.~\ref{fig:RadialPressure}. 
After a numerical transient phase, the pressure rapidly reaches the self-similar regime. Remarkably, this short-time similarity regime holds for almost 4 decades in time and is nicely captured with Wagner theory. Eventually a departure from similarity occurs when $\bar t$ becomes of order 1, and the pressure promptly drops to 0.

\subsection{Normal force induced by drop impact}
\label{sec:normalforce}
Building on the last set of results, we deduce the total net normal force imparted by an impacting drop on the underlying substrate at early times. Integrating the pressure on the wet surface, we have:
\begin{equation}
\tilde F(\tilde  t) = \frac{1}{\sqrt{\tilde t}} \iint_\mathcal{S} \mathcal{P}(\xi,\eta = 0) \, \mathrm{d}S = 2 \pi \sqrt{\tilde t} \int_0^{\sqrt{3}} \frac{3 \xi}{\pi \sqrt{3-\xi^2}} \,\mathrm{d} \xi = 6 \sqrt{3 \tilde t}.
\end{equation} 
The dimensional counterpart of this net total force induced by the drop on the substrate
 therefore reads: 
\begin{equation}
F(t) = 6 \sqrt{3} \rho U^{5/2}R^{3/2} \sqrt{t},
\end{equation}
where the force is seen to increase as $t^{\frac{1}{2}}$ for short times.
Interestingly $F(t)$ could have been inferred directly from energy arguments, with no knowledge of the pressure distribution. Indeed, writing the global kinetic energy conservation for the upper semi-infinite space, we have:
\begin{equation}
\frac{\mathrm d}{\mathrm dt} T = -\oint p \, \boldsymbol u \cdot \boldsymbol n \, \mathrm dS,
\text{ where }
T = \left( \int\!\!\!\int\!\!\!\int \frac{\rho u^2}{2}\, \mathrm dV\right).
\end{equation}
In the context of a flat rising disk, the kinetic energy reduces to $T_\text{disk} = \frac{4}{3}\rho a^3 U^2$ (Lamb \S 102). This expression can immediately be transposed to the impacting drop problem so that $T = 4\sqrt{3} \rho U^{7/2} R^{3/2} t^{3/2}$ (see equation (\ref{eq:kineticenergy}) in \S \ref{sec:expanding_disk}). The power of pressure forces then follows as $\frac{\mathrm d}{\mathrm dt}T = 6 \sqrt{3} \rho U^3{R^2} \left(\frac{Ut}{R} \right)^{1/2}$. Dividing this power by $U$, we recover exactly the previously obtained result for the net normal total force.
This alternate derivation of the normal force provides with yet an other illustration of the relevance of Lamb's analogy for the drop impact problem.


\section{Matching with the viscous solution}
\label{sec:viscous}
 The inertial limit (large Reynolds number hypothesis) investigated so far has allowed us to model the flow within an impacting drop as the winding motion of an inviscid fluid around an expanding disk, appropriately described by an harmonic potential obeying the unsteady Bernoulli equation  (\S \ref{sec:governing_equations}). Actually the agreement between the corresponding theoretical results and numerical Navier-Stokes computations carried out with \gerris\ (encompassing viscous effects) comforted this approximation, see \textit{e.g.} Figs \ref{fig:AxialVelocity} for velocity, \ref{fig:PressionAxiale} for pressure or \ref{fig:dt} for contact line motion comparisons. Most presumably, viscous effects are here dominating only in very thin boundary layers developing along the wet substrate. And indeed, even if the overall agreement between the radial velocity profiles and the inviscid solution is evident, a careful examination of Figure \ref{fig:VitesseRadiale} reveals the presence of these thin layers in the very vicinity of the solid wall. Even if spatially confined, these boundary layers nonetheless play a key role when comes \textit{e.g.} the question of the erosion potential of an impacting drop. Consequently we now set out to describe the inner structure of these viscous layers and to match it to the previously determined outer inviscid solution. Viscous shear stresses and total erosion potential are eventually briefly discussed.
 
\subsection{A simple boundary layer problem?}
Typically, the (inviscid) slip velocity $\tilde{u}_e(\tilde r,\tilde t)=
\frac{2}{\pi} \tilde r/ \sqrt{3 \tilde t -  \tilde r^2}$, here first introduced equation (\ref{eq:slipvelocity}), and the no-slip condition at the substrate, trademark of real fluids, are reconciled through the introduction of a viscous boundary layer.
According to the classic boundary layer theory \citep[\textit{e.g.}][]{Schlichting}, the transverse scale of this layer is $\text{Re}^{-1/2}$, so that an appropriate inner coordinate $\tilde Z$ can be defined via $\tilde{z} = \text{Re}^{-1/2} \tilde Z$. 
The most simple idea at this point is to think that the outer variables scales defined in \S \ref{sec:leading-order} imply that the non-linear terms of the boundary layer equation are negligible when compared to unsteady and viscous terms, so that this equation would simply read:
\begin{equation}
\frac{\partial \tilde{ U}_r}{\partial \tilde{t}} = - \frac{\partial \tilde{p}}{\partial \tilde{r}} +   \frac{\partial^2 \tilde U_r}{\partial  \tilde Z^2},
\end{equation}
where capitalized variables refer to boundary layer quantities.
Considering that Euler equation in the inviscid outer domain reduces to $\frac{\partial \tilde{u}_e}{\partial \tilde{t}} = - \frac{\partial \tilde{p}}{\partial \tilde{r}}$, the boundary layer equation can be recast as the following diffusion equation for the defect velocity:
\begin{equation} \frac{\partial}{\partial \tilde{t}} (\tilde{U}_r-\tilde{u}_e) = \frac{\partial^2 }{\partial  \tilde Z^2} (\tilde{U}_r - \tilde{u}_e). 
\label{eq:diff}
\end{equation}
The corresponding solution can then be expressed as a convolution between the forcing term and the Green function of the heat equation:
\begin{equation}
\tilde{U}_r=\tilde{u}_e(\tilde r,\tilde t)
  -\frac{\tilde{Z}}{2 \sqrt{\pi}} \int_0^{\tilde t} \exp \left(-\frac{\tilde{Z}^2}{4(\tilde t-\tau)} \right) \frac{\tilde u_e(\tilde r,\tau)}{(\tilde t-\tau)^{\frac{3}{2}}} \,\mathrm{d} \tau.
\label{eq:faux}  
\end{equation}
Unfortunately this solution leads to a paradoxical cancelling of shear stresses at the wall. We conjecture that this unreasonable result stems from the fact that the sharp longitudinal variations associated with the contact line motion have here been disregarded. Specifically non linear terms do balance unsteady terms, at least near the contact line location $\tilde r=\sqrt{3 \tilde t}$. As a result, the boundary layer actually grows from this moving point both in space and time.
While a comprehensive analysis of this problem demands a careful balance of each term likely resulting in a non linear boundary layer problem, beyond the scope of the present study, we nonetheless propose in the following an approximation based on an analogy with boundary layers developing behind shockwaves. 

\subsection{Approximation of the drop impact boundary layer via an analogy with shock-induced boundary layers}
\begin{figure}
\centering
\includegraphics{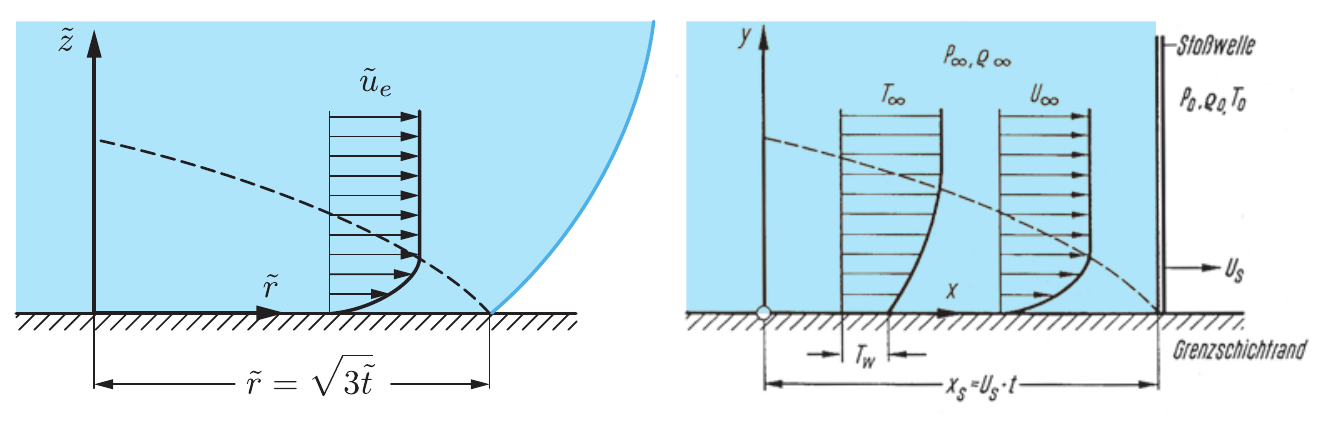}
\caption {Left: Sketch of the contact line during its motion and of the growing boundary layer in its trail, analogous to that developing behind a shockwave. Right: Shockwave-induced boundary layer, reproduced from the german edition of Schlichting textbook \citep{Schlichting}. Notations are from Schlichting, with a correspondence between $x$ and $\tilde r$. Note that in the shockwave case, $U_\infty$ and $U_s$ are both constant.}
\label{fig:Nplus1}
\end{figure}

We now depict qualitatively the inner viscous structure of the velocity field by using a simple analogy.
First remembering the tank-treading movement in the vicinity of the contact line observed and discussed in \S \ref{sec:contactlinenumericalobservations}, we point out the violent change in radial velocity when passing through the contact line. In other words, the contact line embodies a neat discontinuity where the slip velocity sees its value suddenly change from 0 to $\tilde{u}_e$. Building on this observation, we consider in the following the contact line as a kind of shock wave sweeping the substrate, and seeding a boundary layer in its trail (see figure \ref{fig:Nplus1}). 
This problem is classic in compressible flows and was solved by \citet{Mirels55} in the context of a shock tube \citep[see][for more details]{Schlichting}. In this study, a fluid initially at rest is swept by a shockwave travelling at celerity $U_s$ in the direction $x$ and instantly acquires an impulse of velocity $U_\infty$ in the process. Behind the normal shockwave is left a growing viscous boundary layer.

The \textit{Ansatz} for Mirel's solution is to introduce  $\eta_m = z/\sqrt{t - x/U_s}$
as the self-similar variable. This variable not only takes into account time variations but also longitudinal effects from the shock backwards in $x$. 
Disregarding any pressure gradient but considering both unsteady and nonlinear effects, the momentum equation may be rewritten in terms of $\eta_m$ and of the velocity $U_\infty f'(\eta_m)$:
\begin{equation}
f'''(\eta_m) +\frac{1}{2}(\eta_m -\frac{U_\infty}{U_s}f(\eta_m))f''(\eta_m)=0,
\text{ with } f(0)=f'(0)=0, \text{ and } f'(\infty)=1.
\label{eq:schli}
\end{equation}
Note that compressible effects have here been absorbed via an appropriate Lees-Dor\-od\-nit\-syn's transformation \citep[see][]{Stewartson1964}.
Two limiting cases clearly emerge from the picture. For large $U_\infty/U_s$ (and after a rescaling and a change of sign due to the choice of origin), the velocity profile tends to a Blasius profile. Conversely, for small values of the velocity ratio, the velocity rather adopts an error function profile. Note that profiles corresponding to intermediate values of this ratio can be found in Schlichting's textbook.  

\begin{figure}
\centering
\includegraphics[]{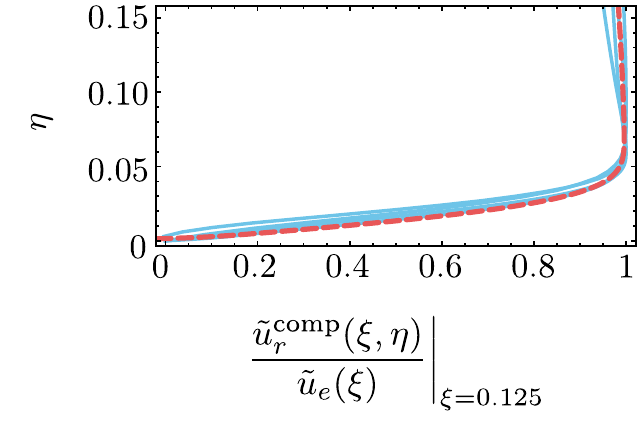}
\includegraphics[]{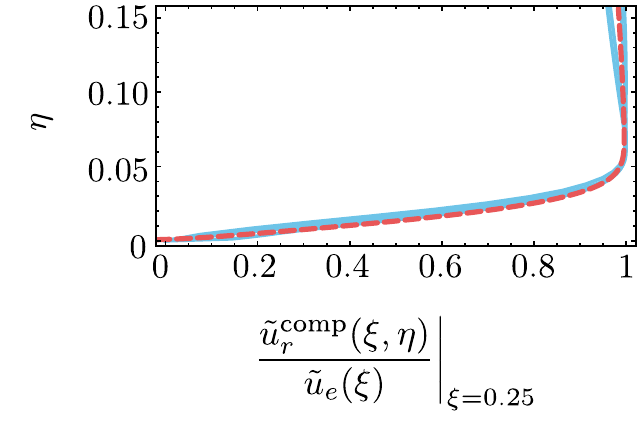}
\includegraphics[]{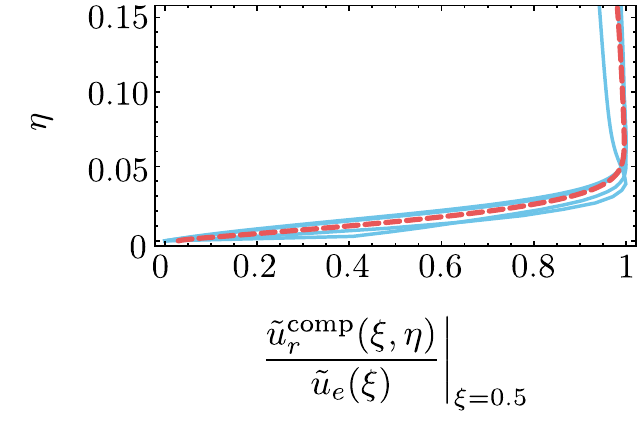}
\includegraphics[]{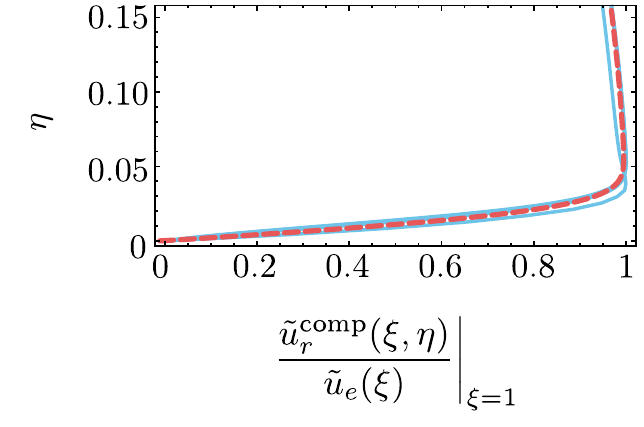}
\includegraphics[]{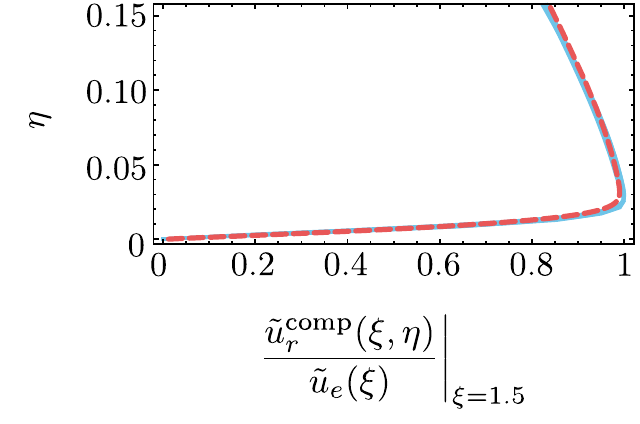}
\caption{Inner-boundary layer radial velocity profiles at different locations $\xi:$ 0.125, 0.25, 0.5, 1 and 1.5.
Blue solid lines correspond to numerical solutions obtained with \gerris\ at $\bar{t} = 5\times 10^{-3}, 10^{-2}, 5\times 10^{-2}$ and $10^{-1}$ for $\text{Re}=5000$ and $\text{We}=250$ and represented in the self-similar space.
Note that velocities are rescaled by their maximum value.
The red dashed lines stand for 
the theoretical composite solution (equation~(\ref{eq:compozit})) blending the self-similar viscous boundary layer solution with the self-similar Wagner inviscid solution for impact.
The composite solution is also rescaled by the edge velocity $\tilde{u}_e(\xi)$ given by equation~(\ref{eq:slipvelocity}).
}
\label{fig:CoucheLimite}
\end{figure}

From this sound result we may by analogy transpose this approach to the drop impact problem (see Fig.~\ref{fig:Nplus1}). Obviously the outer solution for the drop impact problem is quite more complex for neither $U_s$ nor $U_\infty$ are constant. The core idea consists in drawing a parallel between the shock (at position $U_s t$) and the contact line (at position $\sqrt{3 \tilde t}$) on the one hand, and between the steady slip velocity $U_\infty$ and $\tilde u_e(\tilde r,\tilde t)$ on the other hand. Following this simple analogy the longitudinal velocity is approximated with:
\begin{gather}
\label{eqn:inner-sol}
\tilde{ U}_r(\tilde{r},\tilde{z},\tilde{t}) = \frac{2 \tilde{r}}{\pi \sqrt{3 \tilde{t} - \tilde{r}^2}} \; f' \left( \frac{\tilde{z}}{2 \sqrt{\tilde{t}-\tilde{r}^2/3}} \sqrt{\text{Re}} \right).
\end{gather}
where $f'$ is solution of an equation which is analogous to Eq. (\ref{eq:schli}).
The so-called composite solution \citep{VanDyke1975}, which is an expansion valid in the ideal fluid and in the boundary layer, then follows:
\begin{equation}
\begin{split}
\tilde u^{\text{comp}}_r = &-\frac{2}{\pi} \int_0^\infty \frac{ \sqrt{3}k \cos (\sqrt{3} k)-\sin(\sqrt{3}k) }{k}e^{-k \frac{\tilde{z}}{\sqrt{\tilde t}}} J_1(\frac{k\tilde{r}}{\sqrt{\tilde t}} )\, \mathrm dk + \\
&+\frac{2 \tilde{r}}{\pi \sqrt{3 \tilde{t} - \tilde{r}^2}} \left( \; f' \left( \frac{\tilde{z}}{2 \sqrt{\tilde{t}-\tilde{r}^2/3}} \sqrt{\text{ Re}} \right) - 1 \right).
\label{eq:compozit}
\end{split}
\end{equation} 
In practice we approximated $f'$ with erf function. Figure~\ref{fig:CoucheLimite} proposes a comparison between the numerical velocity profiles extracted from \gerris\ computations and this approximation, which proves to provide a fairly good description for the flow. 
As a side note, we remark that replacing the error function with Blasius profile yields a slightly more marked deviation between theory and numerical results. That said we chose not to tune the velocity ratio appearing in equation~(\ref{eq:schli}) as \textit{(i)} this is too speculative and \textit{(ii)} such adjustment is certainly beyond the limits of our analogy.  

It is interesting to note that for $\xi$ smaller than $\sqrt{3}$, Mirel's self-similar variable $\eta_m$ tends to $\text{Re}^{1/2} \eta = \tilde Z / \sqrt{\tilde t}$, so that the vertical structure for solution (\ref{eqn:inner-sol}) now simply involves $\mathrm{erf}\left(\frac{1}{2}\text{Re}^{1/2} \eta\right)$. Actually, this solution is merely the purely diffusive solution of equation~(\ref{eq:diff}) for constant forcing ($\tilde u_e$ constant in time). 

\begin{figure}
\centering
\includegraphics[]{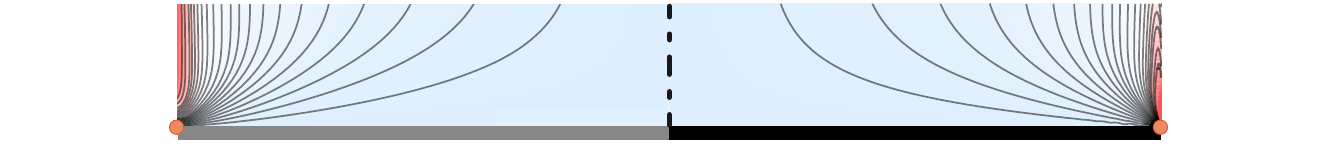}
\caption {Left: Isolines for the radial velocity $\tilde u_r$ extracted from the numerical simulations.
Right: Theoretical isolines for the radial velocity given by the composite expansion $\tilde{u}^{\text{comp}}_{r}$. Note that the transverse scale has here been stretched to visualize the boundary layer.
}
\label{fig:Nplus2}
\end{figure}

From the previous results we may extract several quantities, such as the displacement thickness or the locus of iso-velocities. The displacement thickness $\delta_1$ can readily be estimated with $f'=\text{erf}$ as: 
\begin{equation}
\delta_1 = \frac{1}{\sqrt{\text{Re}}}\int_0^{+\infty} \left( 1 - \frac{\tilde{U}_r}{\tilde{u}_e} \right) \text{d} \tilde{Z} = \frac{2}{\sqrt{\pi } \sqrt{\text{Re}} }   \sqrt{\tilde{t}-\tilde{r}^2/3}.
\end{equation}
Similarly, isolines for the velocity can be extracted both for \gerris\ computations and for boundary layer theory. Figure~\ref{fig:Nplus2} provides with a qualitative comparison between theory and numerical results, and it can be remarked that the overall prediction is more than just qualitative.

Though the velocity ratio $(2 r/\pi /\sqrt{3 \tilde t - \tilde r^2} )/(\sqrt{3 \tilde t}/2 /\tilde t )$ (counterpart of $U_\infty/U_s$ in equation~(\ref{eq:schli})) is infinite near the shock, we note that the agreement between numerical and theoretical solutions is actually surprisingly good. We conjecture that whenever this ratio decreases to a value lower than one, \textit{i.e.} near the centre and for large times where this ratio behaves as $(4 \tilde r)/(3\pi)$ so tends to 0, the error function approximation emerges as the solution of equation~(\ref{eq:schli}).
Eventually we remark that an in-depth analysis of these phenomena demands a more involved description for the boundary layer \citep[such as the Interactive Boundary Layer theory, see \textit{e.g.}][]{LagreeCISM} and/or a deeper analysis of the Wagner region \citep{Oliver2002,Korobkin2007,Oliver2007}.

\subsection{Estimation of the shear stress and the total drag}
\begin{figure}
\centering
\includegraphics[width=6cm]{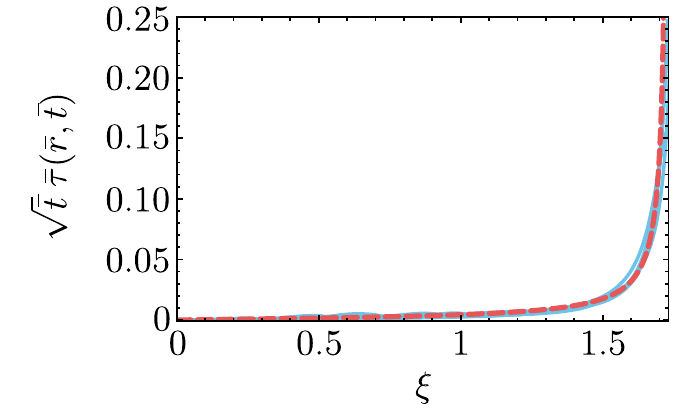}
\includegraphics[width=6cm]{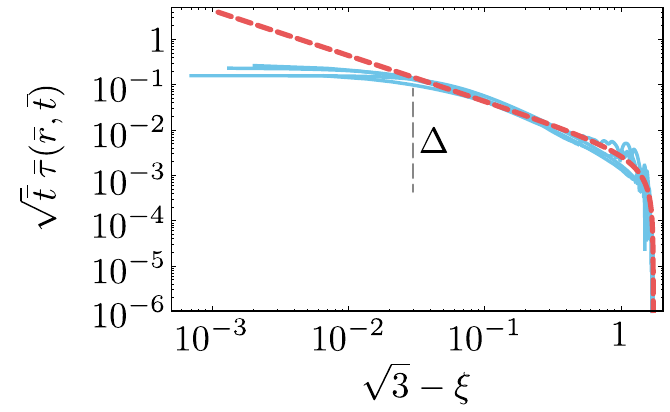}
\caption{Left: Numerical and theoretical shear stress distribution underneath the drop, represented in the self-similar space 
 (the numerical data are taken at times $\bar t =$ 5$\times 10^{-3}$, $10^{-2}$, 5$\times 10^{-2}$ and $10^{-1}$). Right: Same data represented as a function of the distance from the contact line (log plot). This representation reveals a cut-off distance $\Delta$ from which the $1/x$ singularity is screened.
Importantly the numerical mesh size has been chosen to be small enough ($\Delta x = 5\times 10^{-4}$) to ensure the resolution of the fine-scale motion in the vicinity of the contact line. 
}
\label{fig:shear}
\end{figure}

With this boundary layer solution, we are now in a position to provide with an estimation of the wall shear stress
$\tilde \tau = \left.\frac{1}{\text{Re}} \partial \tilde{U}_r/\partial \tilde Z \right \vert_{\tilde Z = 0}$, \textit{i.e.} the viscous component of the stress which has been  disregarded so far.
And indeed this quantity is of paramount importance as far as raindrop-induced erosion of erodible beds is concerned  \citep{Ellison1945,Rein1993,Lagree2003,Leguedois2005}. 
Upon using equation~(\ref{eqn:inner-sol}) (with the same erf approximation for function $f'$ as before), we readily obtain:
\begin{equation}
\tilde \tau(\tilde r,\tilde t) =  \frac{2 \sqrt{3} \tilde r}{\pi^{\frac{3}{2}} \text{Re}^{1/2}  (3 \tilde t - \tilde r^2)}.
\end{equation}
This theoretical prediction is confronted Fig.~\ref{fig:shear} with numerical profiles for the shear stress extracted from \gerris\ computations, and is shown to nicely agree with observations.
From this local distribution for the stress we may infer the total drag induced with a drop impact, by integration over the wet area:
\begin{equation}
\tilde D(\tilde t) = \int_0^{2 \pi} \!\!\!\int_0^{\sqrt{3t}} \tilde \tau(\tilde t) \; \tilde r \; \text{d}\tilde r \; \text{d} \theta.
\end{equation}
Unfortunately this integral diverges due of the $1/x$ singularity developing in the near contact line region, and visible from Fig.~\ref{fig:shear} left. Such singularities are usually a signature of an additional physics in the diverging region, not taken into account in the model. And indeed, Fig.~\ref{fig:shear} right reveals that the calculated shear stress significantly deviates from the theoretical prediction at some small distance $\Delta$ from the contact line position to reach a maximum value. Now integrating the local shear stress up to $\tilde r = \sqrt{3 \tilde t} - \Delta$, where $\Delta$ is this small cut-off length, we can provide an estimation for the drag at leading order in $\log(\Delta)$: 
\begin{equation}
\tilde D(\tilde t) = 
 3 \sqrt{\frac{\tilde t}{\pi \text{Re}}}  \left(-2 \log
   \left(\frac{\Delta}{\sqrt{\tilde t}}\right)-4+\log (12)\right).
\end{equation}
Upon noting that this quantity can be dimensionalised with $\rho U^2 R^2$,  
the expression for the total drag in dimensioned variables follows:
\begin{equation}
D(t) = \frac{3}{\sqrt{\pi}} \mu^{\frac{1}{2}} \rho^{\frac{1}{2}} U^2 R \sqrt{t} \left( -2 \log \left(\frac{\Delta}{\sqrt{\frac{Ut}{R}}} \right)-4+\log(12) \right).
\end{equation}
Noticeably, the departure from the theoretical prediction pinpointed out in Fig.~\ref{fig:shear} right seems to occur at a precise location in self-similar variables, therefore suggesting a $\sqrt{\tilde t}$ time dependence for $\Delta$. From the numerical computations the value of $\Delta/\sqrt{\tilde t}$ can be estimated to be around $0.03$. Note that this is obviously a crude estimation, which nonetheless allows to propose the following estimate for the impact-induced drag:
\begin{equation}
D(t) \simeq  10.7  \mu^{\frac{1}{2}} \rho^{\frac{1}{2}} U^2 R \sqrt{t}.
\end{equation}

To further refine this prediction, the true nature of the cut-off length $\Delta$ needs to be clearly identified.  
Several candidates for governing this quantity naturally emerge, with for example the viscous $1/\text{Re}$ regularisation length in the vicinity of the contact line region or the inertial matching with the Wagner inner layer of typical size $(d(t)/R)^2$). 
This requires further investigation.

\section{Further comments and conclusion}
\noindent Capillary phenomena as well as possible aerodynamic effects from the surrounding gas have been disregarded so far. In this last part we shall estimate their influence on impact and discuss natural extensions of the present work. A summary and general conclusion then follow in \S \ref{sec:conclusion}.
\subsection{Influence of capillary phenomena}
The Weber number provides with a global measure of the ratio of available kinetic energy to surface energy. For low values of this number (with respect to unity), drops typically bounce \citep{Richard2000} while preserving their shape or gently spread \citep{PasandidehFard1996}
according to the wetting properties of the underlying substrate. Note that even in this regime of droplet deposition, fast phenomena associated with the imbalance of surface stresses can set in \citep{Stebnovskii1979}.
When the initial kinetic energy of the drop can no more be neglected in front of the surface energy (We~$\sim$~1), a surface wavefield starts to develop on the drop, shaping it into a characteristic liquid pyramid or torus \citep{Renardy2003}. For even higher values of the Weber number, such as the inertial limit We $\gg$ 1 investigated in the present paper, we do not expect capillary phenomena to have a significant influence on a global scale, but locally surface tension can still play a dominant role. For example, the high-curvature turnaround region at the lamella root is typically a place where capillarity presumably plays an important role. But due to scale separation, this region is invisible at our level of description. 
Indeed, in classic impact analyses  see \textit{e.g.} \cite{Oliver2002}, the typical extent of this intermediate Wagner region associated with highly curved interfaces is found to be $O(\varepsilon^2)$, to be compared both with the $O(\varepsilon)$ size of the main impact region considered throughout this paper (see~\S\ref{sec:leading-order}) and with the $O(\varepsilon^3)$ thickness of the lamella. In the framework of our first-order theory, we therefore do not anticipate appreciable deviations stemming from this zone. Conversely, for a correct description of the ejected liquid sheet feeding conditions and of the pressure fall-off near the lamella root reported Fig.~\ref{fig:RadialPressure}, an accurate representation of this matching region appears mandatory. 

\noindent The contact line is an other region where marked effects from capillarity are to be expected. Drop impact is characterized with fast motions near the contact line. This violent dynamic wetting phenomenon can arguably bring about issues in our description of impact. Actually \citet{Blake1999} demonstrated that nonlocal hydrodynamics could play a significant role in the dynamic contact angle selection. Based on experimental data, \citeauthor{Blake1999} further put forward the possible `mutual interdependence' between the phenomena in the near contact line region and the far-field hydrodynamics. This complex interplay was further confirmed in the context of drop impact by \citet{Sikalo2005}, but especially for the late receding phase. Interestingly, these authors demonstrated that the early evolution of the contact angle was quite insensitive to the experimental conditions and fairly well captured by the contact angle of a truncated sphere. This nice agreement certainly advocates for a predominance of inertial effects over capillary corrections emanating from the dynamic contact line, at least in the early stage of impact. And indeed, remembering that shortly after impact the fluid motion in the contact area is essentially vertical, it appears likely that the point of contact can be determined with mere inertial arguments. In our simulations, dynamic effects have been disregarded in the description of the contact angle, which has been set to the constant value $\pi/2$. The agreement between our simulations and the purely inertial theory is again an indication of the unimportance of dynamic wetting. It might further be interesting to note that the surface energy gained by wetting the solid is of the order of $1/$We when rescaled by the initial kinetic energy. Again this heuristically rules out any leading effect from wetting in the short-term dynamics. This ratio evolves with time though, and ultimately wetting phenomena become dominant, as evidenced by the late $t^{1/10}$ spreading behaviour consistent with Tanner's law in the experiments of \citet{Rioboo2002}. 


\subsection{Influence of ambient air}
\label{sec:air}
For about a decade or so, there has been an increasing realization of the role played by surrounding air in liquid impact in general, and drop impact in particular. Following key experiments performed by \citet{Xu2005} on air-induced splash triggering, a number of studies have focused on the events preluding liquid sheet ejection. The first significant effect of surrounding air is to impart a dimple-like deformation in the bottommost region of the drop (see experimental observations of the dimple obtained by \citealt{Thoroddsen2005} and X-ray ultra-fast imagery of the complex dynamics of this air pocket by \citealt{Lee2012}). \citet{Smith2003} first depicted theoretically this process by coupling lubrication in the squeezed air film and potential flow inside the drop. These authors notably evidenced the presence of off-axis pressure peaks. While more recent studies raised doubt about the link between this dimple formation and splash triggering per se -- that might merely be a secondary independent consequence of the presence of surrounding gas \citep{Duchemin2011}, this gas pocket is nonetheless formed over timescales and lengthscales overlapping that of the phenomenon reported in the present paper \citep{Mani2010}. It is therefore legitimate to question about the impact of this air entrapment phenomenon on our results. 

\noindent In order to investigate these effects, we performed a simulation of a liquid drop approaching a solid substrate, deforming as a result of lubrication pressure rise in the film, and finally impacting the substrate. \begin{figure}
\centering
\includegraphics[]{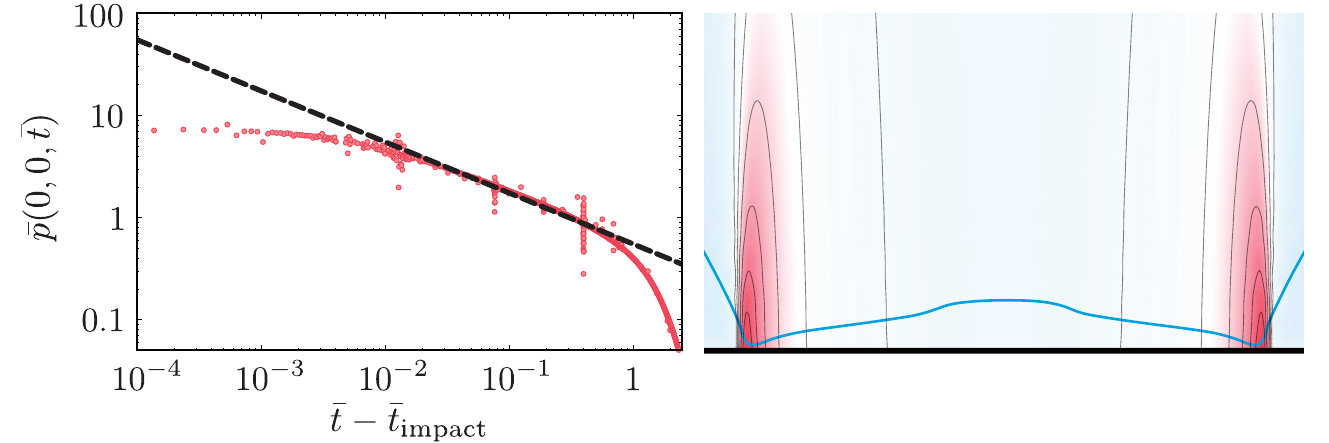}
\caption{Left: Time evolution of the pressure as measured under an impacting drop with air-induced dimple formation (\textit{i.e.} bubble entrapment) taken into account. The red trace monitors the pressure at the origin. The physical parameters of this \gerris\ simulation are $\text{Re}=5000$ and $\text{We}=250$. The superimposed black dashed line corresponds to  the theoretical solution $\bar{p}(0,0,\bar{t})=\frac{\sqrt{3}}{\pi} (\bar{t}-\bar{t}_\text{impact})^{-\frac{1}{2}}$ delayed from $\bar t_\text{impact}$, where $\bar{t}_\text{impact}$ is the real impact time (in the numerical simulations, $\bar{t}_\text{impact}$ corresponds to the time at which liquid and solid are just one grid cell away). Right: Close-up of the bottommost point of the drop in the numerical simulation (at $\bar t = 1.54\times 10^{-1}$). The position of the interface is materialised with a blue line. The colormap illustrates the distribution of the pressure field within the drop and in the gas layer. Noteworthy enough the isopressure lines seamlessly cross the interface, revealing the transparency of the dimple to pressure. Note that for the sake of clarity, the vertical scale has here been magnified by a factor 22.}
\label{fig:PressionDimple}
\end{figure}
Figure~\ref{fig:PressionDimple} represents the time evolution of the pressure exerted on the support at the axis. Considering that air delays the moment of impact \citep{Mani2010}, we introduce a time shift $t_\text{impact}$ corresponding to the moment where the drop and the solid are only a grid cell apart. Interestingly our results reveal that the pressure at the origin (measuring now the entrapped bubble pressure) is fairly well captured by relation~(\ref{eq:PressionOrigine}) after replacing $\tilde t$ with the true time from impact $\tilde t - \tilde t_\text{impact}$, that is:
\begin{equation}
\tilde{p}(0,0,\tilde{t})=\frac{\sqrt{3}}{\pi} (\tilde{t}-\tilde{t}_\text{impact})^{-\frac{1}{2}}.
\end{equation}
This agreement between our prediction and a simulation incorporating air entrapment effects not only validates and extends our results beyond the initial scope of Wagner impact theory (disregarding air effects), but also suggests that the results of the present manuscript correspond to the far-field behaviour of an impacting drop in presence of surrounding gas. This observation outlines the appealing prospect of describing both the dimple geometry and associated dynamical fields  by analytical means.
\subsection{Main results}
In this paper, the short-term dynamics of a drop impacting a rigid substrate has been elucidated. A self-similar solution for the impact-induced flow has in particular been unraveled and matched to a self-similar viscous boundary layer. This solution has been intensively validated with numerical \gerris\ computations, and this constant cross-testing between asymptotic theory and multiphase adaptive flow simulations is one of the key feature of the present approach. In the course of this investigation, several important results have been substantiated. These results allow both for a simple yet accurate qualitative depiction of drop impact along with an in-depth quantitative understanding of this phenomenon. These key results are summarised in the following:
\begin{itemize}
\item A fundamental analogy between the water entry of a solid object (Wagner's original problem) and drop impact exists,
\item During the earliest moments post-impact, the contact line follows a tank-treading motion. There is in particular no contact line sweeping motion,
\item 
The impact-induced flow is concentrated in the contact zone, and the far-field merely corresponds to an undisturbed rigid-body motion reducing to a global free-flight at velocity $U$. There is no global or large-scale drop deformation during impact,
\item The position of the contact line is given by the simple relation 
$d(t) = \sqrt{3 R U t}$. Though simple, this locus does not correspond to the cut radius of a truncated sphere,
\item The wet footprint extent of the drop dictates the size of the impact-induced perturbed flow,
\item There is a consistent analogy between the impact-induced flow within the drop and the flow induced by a flat rising expanding disk (Lamb's analogy),
\item  The impact pressure is to be associated with  
the unsteady Bernoulli contribution $-\partial_t\phi$.  It 
cannot be inferred from usual inertial steady contribution  $-\rho U^2$,
\item As a corollary to the previous point, the impact pressure is extremal at the contact line. It is not maximal at the stagnation point, 
\item A full three-dimensional self-similar solution for the impact-induced flow of an inviscid drop exists and matches quantitatively realistic numerical data on drop impact,
\item Analytical solutions for this flow have been presented in integral forms (with some explicit closed-form expressions along some particular locations), see table \ref{tab:tab1}  (in dimensional form),
\item An original inviscid stagnation point structure with an unexpected 
$r/\sqrt{t}$ slip velocity develops in the vicinity of the origin. The velocity field structure markedly differs from the classic $r/t$ prediction occurring for later times,
\item An approximate self-similar solution for the viscous boundary layer seamlessly match with the inviscid impact-flow (analogy with Mirels shockwave problem),
\item Self-similar variables have the same structure
 $z/\sqrt{t}$ both in the outer region and in the boundary layer,
\item From the knowledge of the distribution of the dynamical fields across the wet area, the expressions for the normal and tangential total force on the substrate are provided,
\item The asymptotic solution is found to be numerically valid over several decades in time. This solution was found to be insensitive to air-induced dimple formation,
\item For times of order one, the present results remain at least qualitative.
\end{itemize}

\begin{table}
\begin{center}
\begin{tabular}{@{}ll@{}}
$(r=0, \quad   z =O(d(t))  >   0^+ )$ &   $ (0 \leq r < d(t)=\sqrt{3 RUt},
\quad z =0^+)$   \\[12pt]
$p(0,z,t)= \frac{3 \sqrt{3} \rho U^{\frac{5}{2}} R^{\frac{3}{2}}}{\pi} \frac{\sqrt{t}}{3 U R t + z^2}$    & $p(r,0^+,t)= \frac{3 \rho U^2 R}{\pi \sqrt{3 U R t - r^2}}$     \\
$u_r(0,z,t) = 0$      &
$u_r(r,0^+,t)= \frac{2 U r}{\pi \sqrt{3 U R t - r^2}}$
 \\
$u_z(0,z,t)= \frac{2U}{\pi} \arctan \left(\frac{\sqrt{3 U R t}}{z} \right) - \frac{\sqrt{3 U^3 R t} \; z}{3 U R t + z^2}-U$   &    $u_z(r,z,t)=0$   \\
$\;$ & $\;$\\
\hline
 $(d(t) \gg r ,\quad d(t) \gg z> 0^+)$ &    $  (0 \leq r < d(t)=\sqrt{3 RUt},\quad z =O(\text{Re}^{-1/2} d(t))$\\[12pt]
 $u_r(r,z,t)=\frac{2}{\pi \sqrt{3}} \; U^{\frac{1}{2}} R^{-\frac{1}{2}} \frac{r}{\sqrt{t}}$  
&  $U_r(r,z,t)= \frac{2 U r}{\pi \sqrt{3 U R t - r^2}} \text{erf}\left(\sqrt{\frac{\rho U R }{\mu (U R t -\frac{r^2}{3})}}\; \frac{z}{2}\right)$    \\
$u_z(r,z,t)=-\frac{4}{\pi \sqrt{3}} \; U^{\frac{1}{2}} R^{-\frac{1}{2}} \frac{z}{\sqrt{t}}$   
&  $\tau(r,t)=\frac{2 \sqrt{3} \rho^{\frac{1}{2}} U^{\frac{3}{2}} R^{\frac{1}{2}} \mu^{\frac{1}{2}}}{\pi^{\frac{3}{2}}} \frac{r}{3 U R t - r^2}$   \\[12pt]
\hline
$p(0,0,t)=\frac{\rho U^{\frac{3}{2}} R^{\frac{1}{2}}}{\pi} \sqrt{\frac{3}{t}}$      $\qquad F(t)= 6 \sqrt{3} \rho U^{\frac{5}{2}} R^{\frac{3}{2}} \sqrt{t}\quad $  & $ 
 D(t) \simeq  10.7  \mu^{\frac{1}{2}} \rho^{\frac{1}{2}} U^2 R \sqrt{t}$\\
 \hline
\end{tabular}
\end{center}
\label{tab:tab1}
\caption{Summary of the main results of the paper in dimensioned form. The top part of the table refers to ideal fluid results (left: closed-form results along the axis of symmetry, right: along the substrate). The left middle part sums up inviscid stagnation point results, and the right middle part summarises the viscous boundary layer results.
Observables such as the net normal force $F(t)$ and tangential force $D(t)$ are reminded as well in the bottom part of the table.
} 
\end{table}

\subsection{Perspectives}
The results obtained in the present manuscript offer several appealing prospects. On the role of surrounding air first, the last results of \S \ref{sec:air} support the idea that the dimple geometry and characteristics could be derived analytically, with a far-field corresponding to the here presented flow. The question of the role of air in suppressing the pressure divergence \citep{Josserand2015}, or in altering the shear stress distribution is also a central point for a correct description of a single drop impact action.
Similarly, the question of the scope of the present results for impact on a liquid film or on a soft/erodible substrate is also of interest. Indeed, the analytical and numerical toolboxes developed here might well be transposed to other rheologies (such as Bingham, see \citet{Staron2013} or granular media, \citep{Lagree2011}).
While the inverse-square-root singularity of the pressure is integrable and yields a finite normal force, the sharper singularity observed in the viscous shear stress distribution leads to a divergence for the total drag. From the simulations, it appears that the singularity is screened over a lengthscale $\Delta$, but the precise nature of this cut-off scale is still unclear. As discussed in \S \ref{sec:viscous} both viscous effects and inertial ones are candidates to explain this regularisation. While the rudimentary description for the viscous boundary layer proposed in the manuscript certainly necessitates a refined analysis, the question of inertial effects at the Wagner region scale is also worth studying. Not only an in-depth investigation of such effects might provide an explanation for the regularisation of the shear stress, but it shall shed light over the onset of lamella formation, which still conceals mysteries. Finally the process responsible for the loss of self-similarity observed at intermediate times is still uncertain: confinement effects arising when the impact-flow lengthscale overlaps with the drop extent, eventual deceleration in the far-field region or contact line geometrical departures from the square-root law are all a priori legitimate to explain the final pressure fall-off, and certainly needs further investigation.

\subsection{To conclude}
\label{sec:conclusion}
Within the numerous limits carefully drawn along this paper, a consistent asymptotic description of the dynamics and geometry of drop impacting a solid surface has been proposed.
The results may simply be summarised through three analogies: Wagner water entry (drop impact being the dual of this problem), Lamb's disk winding flow (that accurately represents the flow induced with the impact) and Mirels shockwave-induced boundary layer (remarkably capturing the boundary layer developing in the contact line's trail). 
The original strategy developed throughout this paper has been to validate those three analogies through a constant confrontation between numerical simulations and asymptotic analysis. Our study revealed that
very powerful state-of-the-art adaptive codes now allow to probe all the dynamic features of realistic violent events such as drop impact, but in the meantime, it also emphasised again how powerful and useful asymptotic analysis is in providing an in-depth understanding of such phenomena and in uncloaking the raw data delivered by the code.
Finally our study brought to light some interesting features and observables (such as the particular stagnation point structure, pressure distribution, contact line motion, viscous total drag force) never observed to date neither in simulations nor in experiments. This certainly arouses the exciting prospect of their unveiling in future experimental studies.

\section{Appendix}
\label{sec:Annex}
\subsection{\gerris\ flow solver}
All the numerical simulations were performed with the open-source code \gerris\ \citep[freely downloadable at \url{http://gfs.sourceforge.net} -- see also][for details]{Popinet2003, Popinet2009,Lagree2011}. \gerris\ is a solver of the incompressible Navier-Stokes equations taking into account multiple phases and surface tension. The code makes use of a finite-volume approach and of a Volume-of-Fluid (VoF) method for an accurate description of the transport of the interfaces between two-phase flows. It also features an adaptive mesh refinement procedure allowing for both a precise description of flows with large scale separation and a reduction of computational costs.
\begin{figure}
\centering
\includegraphics[width=6cm]{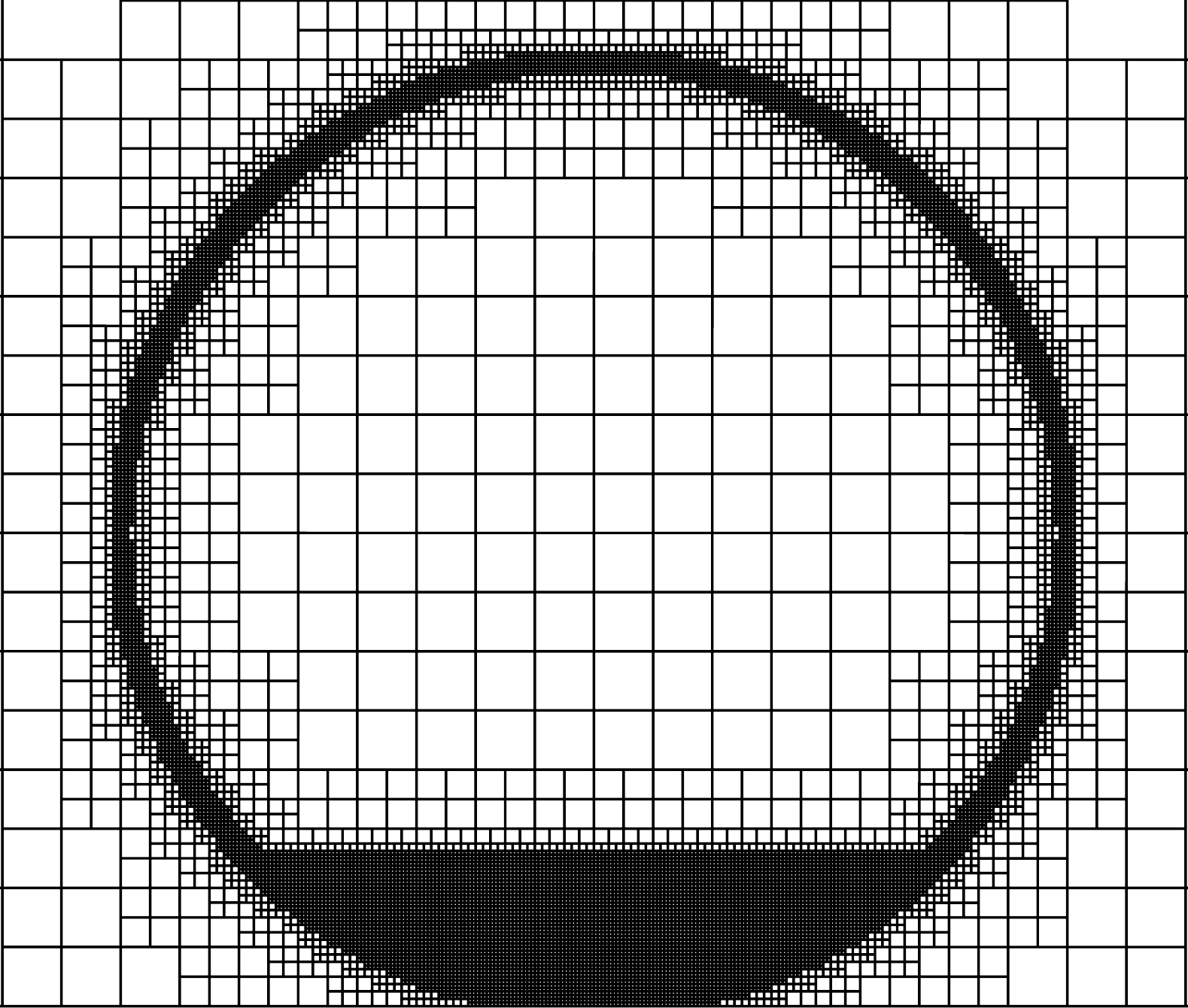}
\caption{Typical mesh structure refined adaptively by \gerris\ flow solver during a simulation.}
\label{fig:mesh}
\end{figure}
Typically in our simulations the finest grid is chosen to be concentrated along free surfaces and within the contact zone to fully capture the features of the pressure field and of the boundary layers (see Fig.~\ref{fig:mesh}). In these areas the corresponding local resolution usually corresponds to 4096$\times$4096 but can reach local density as high as 32768$\times$32768 if needed (examples being Fig.~\ref{fig:tryptic} or Fig.~\ref{fig:radial_pressure}). 

The simulations carried out in this study all correspond to the impact of a water drop in air with a Reynolds number of 5000 and a Weber number of 250. The computations were performed in an axisymmetric configuration. We emphasize that both liquid and air motions were computed with \gerris , but, to be consistent with the post-impact theory developed in this paper the simulations disregarded air cushioning and dimple formation (except explicitly specified, see \S \ref{sec:air}). To avoid dimple formation in this multiphase flow simulation, the initial configuration is set to a slightly truncated liquid sphere already touching the solid surface. The liquid is initialised with a constant downward velocity. The initial sphere penetration $\bar{r}_0 = 10^{-4}$ is at most one grid cell deep (for example the grid spacing is $\Delta x \simeq 5 \times 10^{-4}$ for 4096$\times$4096 simulations). Finally a no-slip boundary condition is enforced at the substrate level. 
\begin{figure}
\centering
\includegraphics[width=6cm]{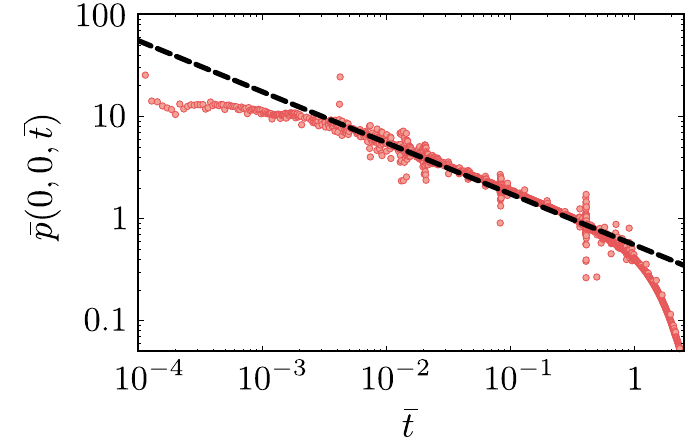}
\includegraphics[width=6cm]{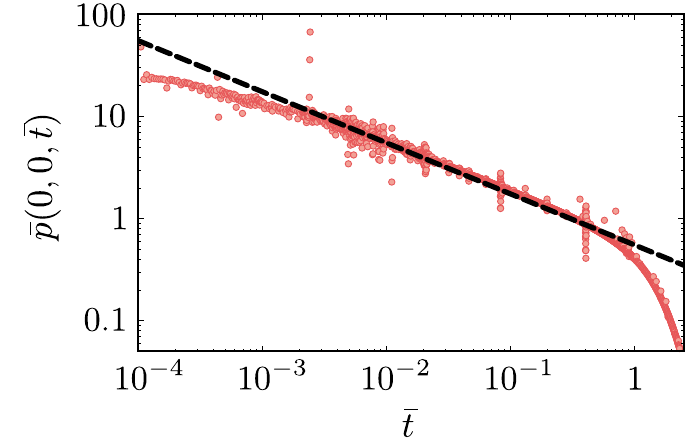}
\caption{Comparison between the pressure measured in the \gerris\ simulations at the origin for two different maximum level of mesh refinement (red dots) and the theoretical prediction (black dashed line). The left panel corresponds to a simulation where the maximal grid density is 2048$\times$2048, and the right panel to a simulation where the maximal density is 4096$\times$4096 (in both cases the physical parameters are $\text{Re} = 5000$ and $\text{We}=250$). The analytical solution of the pressure is given by $\bar{p}(0,0,\bar{t}) = \frac{\sqrt{3}}{\pi} \; \bar{t}^{-\frac{1}{2}}$, see equation~(\ref{eq:PressionOrigine}). After a short transient, both simulations quickly reach the same self-similar asymptotic regime.}
\label{fig:p0mean}
\end{figure}
The reliability of the results has been thoroughly checked with a convergence study on the refinement level, and with particular attention paid for the pressure field and the position of the contact line convergence. Fig.~\ref{fig:p0mean} proposes a comparison between the evolution of the pressure field measured at the origin for two levels of resolution (2048$\times$2048 and 4096$\times$4096). For both cases the numerical solution quickly converges to the theoretical solution $\bar{p}(0,0,\bar{t}) = \frac{\sqrt{3}}{\pi} \; \bar{t}^{-\frac{1}{2}}$ (see equation~(\ref{eq:PressionOrigine})) around $\bar{t} = 5 \times 10^{-3}$  and leaves the self-similar regime at around $\bar{t} = 6 \times 10^{-1}$. 
We remark that after a transient period, both numerical solutions give consistent information and collapse onto the theoretical solution over almost three decades.
Note that the occurrence of sporadic glitches in the numerical solution (see \textit{e.g.} Figs
 \ref{fig:p0}, \ref{fig:p0mean} or \ref{fig:ErreurPression}) are to be related with the classic difficulty of computing the pressure in projection methods, such as the one implemented in \gerris\ \citep{Brown2001,Popinet2003}.
\begin{figure}
\centering
\includegraphics[width=6cm]{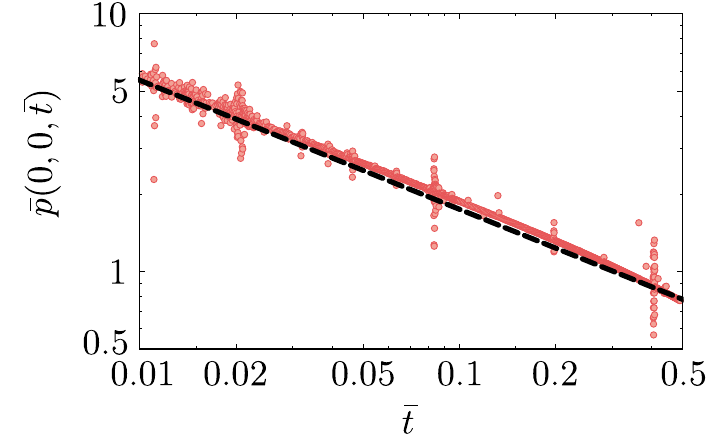}
\caption{Close-up of Fig.~\ref{fig:p0mean} right. Note that the numerical evolution for the pressure is slightly above (about 7 \%) the analytical solution.}
\label{fig:ErreurPression}
\end{figure}
We finally remark that an error of ca. 7 \% between the numerical prediction for the pressure at the origin and the theoretical prediction was consistently noted in our simulations (see Fig.~\ref{fig:ErreurPression}). The nature of this discrepancy is uncertain though, and might either be related to the aforementionned numerical difficulties in computing the pressure or to the limits of our first-order asymptotic description for drop impact.
\subsection{\gerris\ parameter file}
A minimal \gerris\ parameter file allowing to reproduce the results presented in this manuscript is provided here for the reader's convenience (Note for referees and Editors: this section will likely be suppressed in the final version for this script will be on a dedicated example page in \gerris\ website):
\begin{Verbatim}[fontsize=\footnotesize]

Define t0 1e-4
Define Re 5000
Define ReAir 277778
Define We 250
Define VAR(T,min,max)   (min + CLAMP(T,0,1)*(max - min))
Define RHO_EAU 1000.
Define RHO_AIR 1.
Define RHO(T)            VAR(T, RHO_AIR/RHO_EAU, RHO_EAU/RHO_EAU)

4 3 GfsAxi GfsBox GfsGEdge {y=0.5 x=0.5} {
    Time { t=t0 end = 2.5 dtmax = 1e-4 }
    PhysicalParams { L = 2 }
    VariableTracerVOFHeight T
    VariableFiltered T1 T 1
    VariableCurvature K T Kmax
    SourceTension T (1./We) K
    InitFraction T ({
    	double goutte = -(x+t0-1.)*(x+t0-1.) - (y)*(y)  + 1.*1.;
    	return (goutte);
    })
    Init {}{ U=-1*(T) V=0}
# Initial refining
    RefineSurface 12 ( -(x+t0-1.)*(x+t0-1.) - (y)*(y)  + 1.*1. ) 
    Refine ( ((x>0.)&&(x<0.3)&&(y>0.)&&(y<4.)) ? 12 : 4 )
# Refining on tracer gradient
    AdaptGradient { istep = 1 } { maxlevel = 12 cmax = 1e-2 } T   
# Constant value of mesh's level inside the bulk 
    AdaptFunction { istep = 1 } { 
    maxlevel = ( ((x>0.)&&(x<0.3)&&(y>0.)&&(y<4.)) ? 12 : 4 ) 
    cmax = 1e-2 } (T==1)
    RemoveDroplets { istep = 1 } T -1
    PhysicalParams { alpha = 1./RHO(T1) }
    SourceViscosity {  } (T*(1./Re) + (1. - T)*(1./ReAir))  {beta = 1}  
    OutputTime { istep = 1 } stderr
    OutputTiming { step = 0.0001 } stderr
    OutputSimulation { istep = 1 } stdout
    OutputProjectionStats { istep = 1 } stderr
}
#1
GfsBox {
	bottom =Boundary{}
	left = Boundary{ 
		     BcDirichlet U 0 
		     BcDirichlet V 0 }
       }
#2
GfsBox {
	right = Boundary{ 
		     BcDirichlet P 0
		     BcNeumann U 0
		     BcNeumann V 0 }
	bottom = Boundary{ } 
	top = Boundary{  
		     BcDirichlet P 0
		     BcNeumann U 0
		     BcNeumann V 0 }              
       }
#3
GfsBox {
left = Boundary{ 
		     BcDirichlet U 0 
		     BcDirichlet V 0  }	
	 right = Boundary{
	 	     BcDirichlet P 0
		     BcNeumann U 0
	 	     BcNeumann V 0 }            
	    }	    
#4
GfsBox {
	 right = Boundary{
	 	     BcDirichlet P 0
	         BcNeumann U 0
	 	     BcNeumann V 0 }      
	top = Boundary{  
	         BcDirichlet P 0
	         BcNeumann U 0
	         BcNeumann V 0  }     
	left = Boundary{ 
		     BcDirichlet U 0 
	         BcDirichlet V 0 }	 
	   }	
1 2 right
1 3 top
3 4 top
\end{Verbatim}

\subsection{Thanks}
We  warmly thank St\'ephane Popinet, Pascal Ray, Christophe Josserand
and   St\'ephane Zaleski from Institut Jean le Rond $\partial$'Alembert for enlightening discussions on splash phenomena, and Neil Balmforth from the Department of Mathematics of the University of British Columbia for interesting discussions on boundary layers. 

\bibliographystyle{jfm}
\bibliography{self_similar_drop_impact}

\end{document}